# In-memory computing based on all-optically controlled memristor

*Jing Yang, Lingxiang Hu, Liufeng Shen, Jingrui Wang, Peihong Cheng, Huanming Lu,*  Fei Zhuge* and Zhizhen Ye*

J. Yang,[+] L. X. Hu,[+] , L. F. Shen, Dr. J. R. Wang, Prof. P. H. Cheng, Prof. H. M. Lu, Prof. F. Zhuge
Ningbo Institute of Materials Technology and Engineering
Chinese Academy of Sciences
Ningbo 315201, China
E-mail: hmlu@nimte.ac.cn (HL), zhugefei@nimte.ac.cn (FZ)

L. X. Hu, Prof. F. Zhuge
Center of Materials Science and Optoelectronics Engineering
University of Chinese Academy of Sciences
Beijing 100029, China

Prof. F. Zhuge
Center for Excellence in Brain Science and Intelligence Technology
Chinese Academy of Sciences
Shanghai 200072, China

Prof. Z. Z. Ye
State Key Laboratory of Silicon Materials
School of Materials Science and Engineering
Zhejiang University
Hangzhou 310027, China
[+] These authors contributed equally to this work.

**SUMMARY**

Artificial intelligence is widely used in everyday life. However, an insufficient computing efficiency due to the so-called von Neumann bottleneck cannot satisfy the demand for real-time processing of rapidly growing data. Memristive in-memory computing is a promising candidate for highly efficient data processing. However, performance of memristors varies significantly because of microstructure change induced by electric-driven matter migration. Here, we propose an all-optically controlled (AOC) memristor with a simple Au/ZnO/Pt sandwich structure based on a purely electronic tuning mechanism of memconductance. The memconductance can be reversibly tuned only by light irradiation with different wavelengths. The device can be used to perform in-memory computation such as nonvolatile neuromorphic computing and Boolean logic functions. Moreover, no microstructure change is involved during the operation of our AOC memristor which demonstrates superior operation stability. Based on this and its structural simplicity, the device has attractive application prospects for the next generation of computing systems.

## INTRODUCTION

Due to increased use of artificial intelligence (AI) in applications such as face recognition, self-driving vehicles, internet search, Go game and internet of things, increasing demands (especially high speed and low energy) are being placed on the supporting computation hardware. Based on the von Neumann architecture, the central processing unit (CPU) (for data processing) and the main memory (for data storage) of modern computers are physically separated and the data transfer between them consumes most of the time and energy. This is known as the von Neumann bottleneck.[1,2] The reason for this bottleneck is that CPUs are advancing at a much faster pace than memories, causing the computer performance to be increasingly limited by the speed of memories — a phenomenon known as the memory wall.[3]

Unlike the von Neumann architecture, in-memory computing can perform calculations *in situ*, *i.e.*, computations can be carried out in the memory.[4–6] Hence, the latter approach seems promising to eliminate the von Neumann bottleneck or the memory wall. The concept of in-memory computing was pioneered by Kautz in the late 1960s.[7] He proposed cellular logic-in-memory arrays where a computational memory cell is composed of several electronic elements such as transistors, diodes and resistors. Since the late 2000s, due to the slowing of Moore's law, wide applications of AI, and rapid development of nonvolatile memories, in-memory computing, which can be performed in a single electronic element, has been receiving increasing attention. These nonvolatile memories include floating gate transistors,[8–10] memristors,[11–27] phase change memories,[28,29] magnetoresistive memories,[30,31] and ferroelectric memories.[32,33] As illustrated in **Figure 1**, in-memory computing can generally be classified into three categories: nonvolatile neuromorphic computing,[9,13–17,30,31,33] logic-in-memory,[10,34–39] and arithmetic-in-memory.[28,40] Note that nonvolatile neuromorphic computing includes matrix–vector multiplication (according to

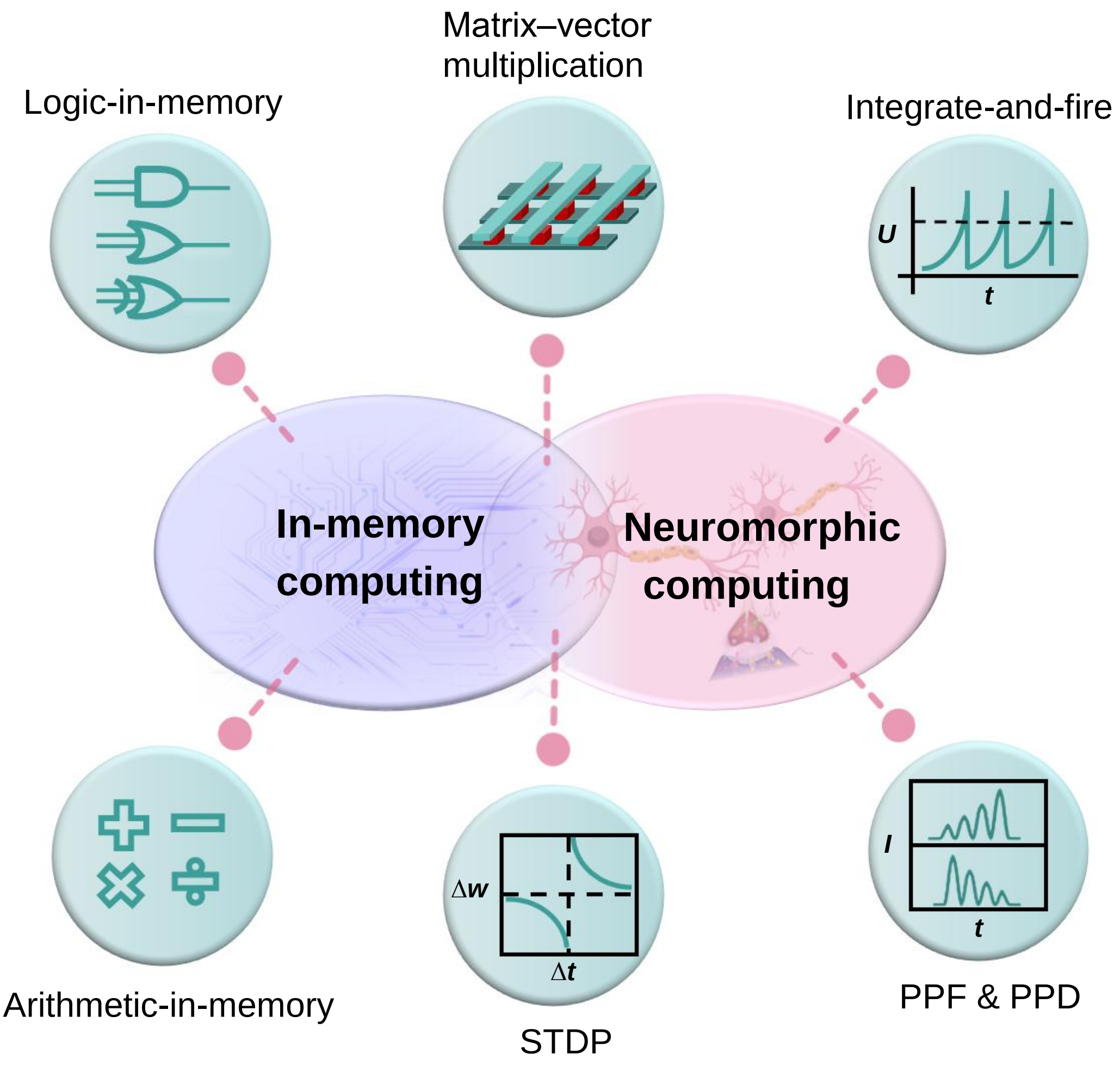

**Figure 1. Different types of in-memory computing and neuromorphic computing.** The overlapped portion includes matrix–vector multiplication and STDP. PPF and PPD denote two types of synaptic short-term plasticity, *i.e.*, paired-pulse facilitation and depression.

universal circuit laws such as Ohm's law and Kirchhoff's law)[9,41–45] and spike-timing-dependent plasticity (STDP).[9,13–16,30,31,46]

Among nonvolatile memories, memristors are considered as one of the ideal candidates for in-memory computing due to their simple two-terminal structure, high operation speed, and low energy consumption.[4–6] Generally, memristors are controlled (in full or in part) by electrical stimuli; that is, memconductance is tuned via electric-field-driven or current-driven ion (or atom) migration within the active layer or at the electrode/active layer interface,[47,48] resulting in microstructure change. However, the microstructure change induces significant cycle-to-cycle and device-to-device performance variations, which are a major obstacle to their practical application.[49,50]

In this work, we propose an all-optically controlled (AOC) memristor which can be operated at low light power densities. This device consists of a simple Au/ZnO/Pt structure. It demonstrates nonvolatile memconductance switching due to light-induced electron trapping and detrapping at intrinsic defects in the oxide, thereby no microstructure change occurs. This device shows excellent operation stability and is used to realize in-memory computing such as nonvolatile neuromorphic computing and logic-in-memory. Therefore, this AOC memristor has a great application potential for designing highly efficient computing architectures.

## RESULTS AND DISCUSSION

### Memristive Switching Behavior

We used polycrystalline ZnO with a wide band gap of 3.2 eV as the active layer (see **Figure S1**). ZnO fabricated by various methods has been widely used as the active layer of memristive devices.[51–54] Herein, the ZnO thin film was deposited at room temperature via RF magnetron sputtering in pure Ar gas. The device demonstrated a nonpolar memristive switching behavior, *i.e.*, from a high memconductance state (HMS) to a low memconductance state (LMS) when measured in the dark (**Figure S2a**). To investigate the switching mechanism, the metal electrodes were replaced by a transparent conducting oxide (Sn-doped $In_2O_3$ (ITO)), which formed quasi-Ohmic contact with ZnO given that the electron affinity of ZnO (4.2 eV) and the work function of ITO (4.3 eV) have similar values.[55,56] No memristive switching phenomenon was observed in ITO/ZnO/ITO (**Figure S2b**). This meant that the switching occurred at the Au/ZnO or the ZnO/Pt interface. It is known that ZnO is naturally an n-type semiconductor because of the existence of intrinsic donor defects, such as oxygen vacancies ($V_O$s) and zinc interstitials ($Zn_i$s).[15,57] Herein, for the sake of simplicity, we consider only $V_O$s. The electron affinity of ZnO was lower than the work functions of Au (5.1 eV) and Pt (5.65 eV), resulting in two back-to-back Schottky junctions at Au/ZnO and ZnO/Pt interfaces (**Figure S3**). When applying a bias voltage, one of the junctions was under reverse bias, whereas the other was forward biased. The memconductance was always dominated by the reverse-biased junction.[56] For example, under a positive voltage, the ZnO/Pt and Au/ZnO Schottky junctions were reverse and forward biased, respectively; the ZnO/Pt junction dominated the electrical properties. Width of the Schottky junction influenced the tunneling current and in turn determined the device memconductance.[40,58] The junction width was determined by the density of ionized oxygen vacancies,[58] which were mainly $V_O^{2+}$s as $V_O^{+}$s are unstable;[59] that is, a higher/lower density of $V_O^{2+}$s led to a narrower/wider width. In an equilibrium state without bias voltage, the numbers of electrons tunneling through the junction along both directions were equal and the net current was zero (**Figure S2c**). When applying a positive bias, there was a flow of electrons into ZnO (**Figure S2d**). Part of the electrons were trapped by $V_O^{2+}$s, which then transformed into neutral $V_O$s. A decrease in the number of $V_O^{2+}$s resulted in an increase in the junction width, thus lowering the memconductance. It deserves mention that the voltage bias induced electrons from the Pt electrode into ZnO, and at the same time, the same number of electrons flowed out of ZnO via the Au electrode. This implies that there were no additional electrons in the ZnO layer or in the Au/ZnO/Pt device. The electrons captured by $V_O^{2+}$s could be considered to be the free electrons already present in ZnO. Therefore, it follows that the density of free electrons in ZnO decreased, which also contributed to the lowered memconductance.

**Optical SET and RESET Behaviors**

The Au/ZnO structure has a mean transmittance of > 60% for light wavelengths from 350 to 1000 nm (**Figure S4**). This suggests that its memconductance could be modulated via light irradiation. Light was injected into the device through the top electrode (Au). Upon illumination with short-wavelength lights (350 and 420 nm), the device current instantaneously increased and reached to a saturation value (**Figure 2a,** top and middle panels). After illumination, the device showed a persistent photocurrent (PPC) phenomenon (see the black curve). Green light (530 nm) irradiation caused a gradual increase in the current followed by a PPC (**Figure 2a,** bottom panel, black curve). PPC is known to be an intrinsic phenomenon in most optoelectronic devices, which can be understood as follows: irradiation with light of appropriate wavelengths results in an increase in the current; after irradiation, although a current decay occurs, the device cannot be restored to its initial conductance state before irradiation, *i.e.*, the conductance state after irradiation is nonvolatile. Therefore, relatively short-wavelength light (350, 420, and 530 nm) could switch the device from LMS to HMS, referred to as the SET operation. On the other hand, when the device was irradiated with long-wavelength light (650, 725, and 800 nm), very weak or even no current change was observed (**Figure S5**).

After irradiating the device with 350 and 420 nm light, subsequent irradiation with 530 nm light resulted in an increase in the photocurrent, whereas 650, 725 and 800 nm light gave rise to a decrease in the current, compared to the case without subsequent irradiation (**Figure 2a,** top and middle panels). Similarly, for the device irradiated with 530 nm light, subsequent irradiation with relatively long-wavelength light induced a decrease in the current (**Figure 2a,** bottom panel). These observations indicate that long-wavelength light could be used to switch the device from HMS to LMS, which is called the RESET operation. To demonstrate the RESET behavior more clearly, the device was first set to HMSs with 350, 420, and 530 nm light, followed by irradiating it with light of long wavelengths 30 minutes after the initial short-wavelength light illumination (**Figure 2b**). The purpose of the 30-minute wait was to ensure that the HMSs were relatively stable. A clear decrease in current could be observed upon the 650, 725 and 800 nm light exposures.

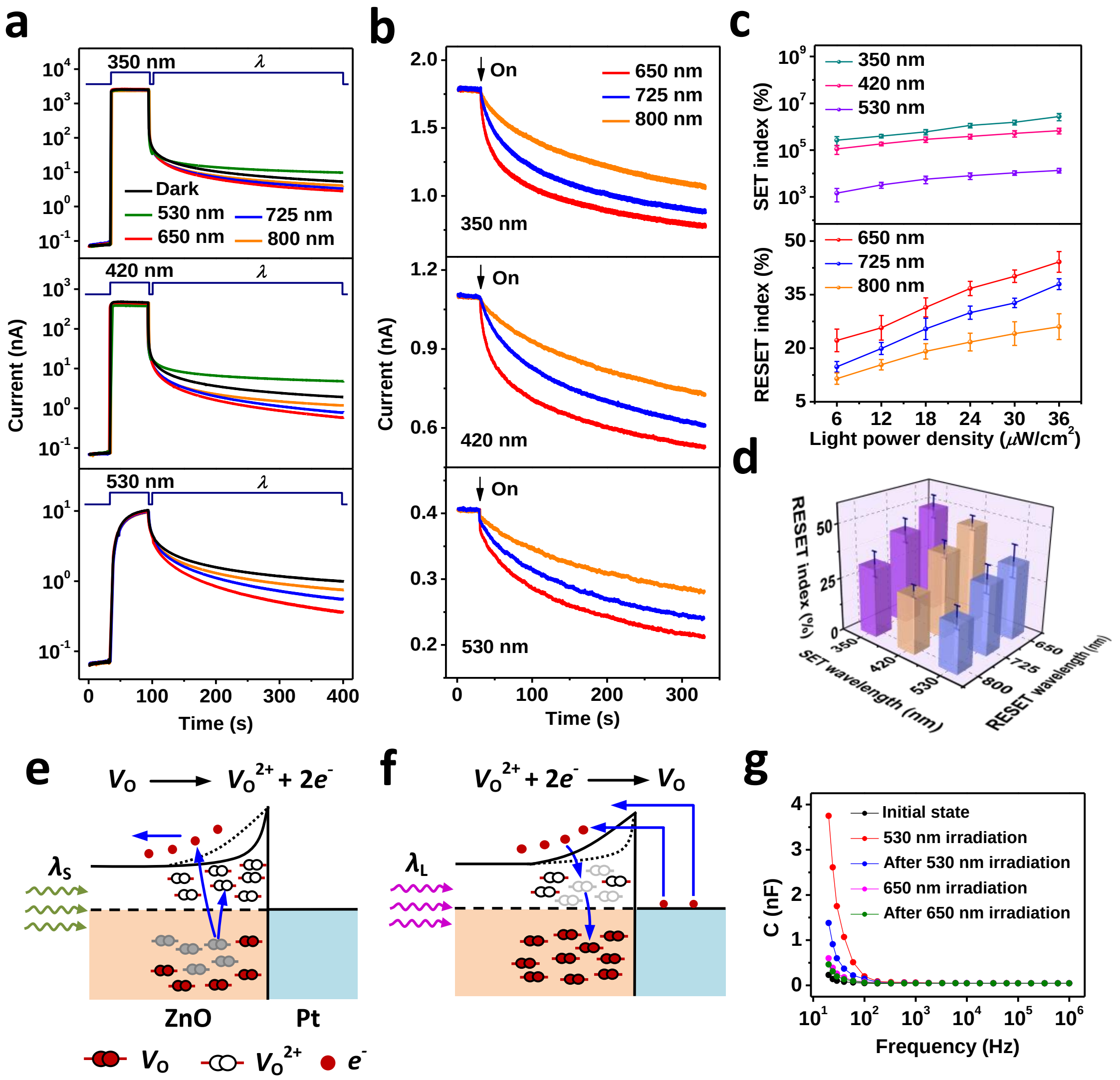

**Figure 2. Optical SET and RESET behaviors.** a) Dependence of the photocurrent response on the wavelength ($\lambda$) of the irradiated light. The device was first irradiated with 350, 420, and 530 nm light (duration ($D$) = 60 s). b) Optical RESET behavior upon exposure to light of various wavelengths ($D$ = 300 s). The device was first set to HMSs by irradiating it with 350, 420, and 530 nm light ($D$ = 60 s). c) Dependence of the SET (top panel) and RESET (bottom panel) indexes on light power density. For the RESET operation, the device was first set to an HMS with 530 nm light, followed by irradiating it with light of long wavelengths 30 minutes after the initial short-wavelength light illumination. d) Dependence of the RESET index on wavelength of the light used for the initial SET and the subsequent RESET operations. e) Equilibrium energy band diagram of the ZnO/Pt Schottky junction after irradiation with short-wavelength light ($\lambda_S$). The $V_O$ ionization reaction is also schematically illustrated (blue arrows). The black dotted line indicates the positions of $E_C$ before irradiation. f) Equilibrium energy band diagram after irradiation with long wavelength light ($\lambda_L$). The electron tunneling and jumping processes as well as the subsequent $V_O^{2+}$ neutralization reaction are also schematically illustrated (blue arrows). The black dotted line indicates the positions of $E_C$ before irradiation. g) Capacitance–frequency characteristics of the AOC memristor. The device was first at the initial LMS (black curve), and then exposed to 530 nm light (red curve), and afterwards kept in the dark for 30 minutes (blue curve); subsequently, the device was exposed to 650 nm light (pink curve), and finally kept in the dark for 30 minutes (green curve). In (a-d, g), the light power densities were maintained at 36 $\mu$W/cm$^2$ and the current values were measured at 10 mV. Note that both the $V_O$ ionization and $V_O^{2+}$ neutralization reactions and electron motion actually occurred under nonequilibrium conditions.

SET and RESET indexes were used to quantitatively study the optical SET and RESET efficiencies. The index values were calculated by the formula ($|I_1 - I_2|/I_1$) × 100%, where $I_1$ and $I_2$ are the device currents before and after irradiation, respectively. As demonstrated in **Figure 2c**, both the SET and RESET indexes increased with increasing light power density. **Figure 2d** illustrates the dependence of RESET indexes on the light wavelength for the initial SET operation and the following RESET operation. Evidently, the RESET index increased with a decrease in wavelength of the light for both SET and RESET.

As mentioned above, the ZnO/Pt Schottky junction dominated the device memconductance under a positive read voltage (10 mV). It could be deduced that this junction played a key role in the optical SET operation (**Figure 2e**). Specifically, light induced an ionization reaction in which neutral $V_O$ transformed into $V_O^{2+}$s. An increase in the number of $V_O^{2+}$s caused a decrease in the junction width, thus leading to an increased memconductance. The nonvolatility of the light-induced memconductance states is likely due to the following reasons: i) the free electrons generated in the Schottky barrier region were pulled into ZnO by the built-in electric field and therefore could not recombine with $V_O^{2+}$s, and ii) an energy barrier originating from the outward relaxation of bonds around the oxygen vacancy sites must be overcome to neutralize $V_O^{2+}$s.[60] The visible light (*e.g.*, 530 nm) response of such wide band gap ZnO is due to an abundance of $V_O$s with a wide distribution of energy levels given that ZnO was deposited in pure Ar.[22,61] It is also supported by photoluminescence measurement (**Figure S6**), in which broad deep emission bands ranging from 480 to 900 nm was observed. In addition, it has been reported that $V_O$s in ZnO could have energy levels ranging from 0.2 to 1.3 eV below the conduction band minimum.[62]

Contrary to the optical SET, an optical RESET is expected to result from widening of the Schottky junction due to a reduced density of $V_O^{2+}$. As mentioned previously, no significant photocurrent was generated by irradiation with 650, 725, and 800 nm light for the device in LMS (**Figure S5**). Therefore, we can deduce that for the device after the optical SET, *i.e.*, in an HMS, ionization of $V_O$s could be ignored under irradiation with such long-wavelength light given a lower density of $V_O$s in ZnO compared to the case without the optical SET (in LMS). It has been reported that in a metal/oxide/metal junction,[63,64] electrons in metal could be injected into the oxide via internal photoemission or photoassisted tunneling during light irradiation. Hence, possible explanation for the optical RESET is that electrons in the Pt electrode entered the conduction band of ZnO upon long-wavelength light irradiation, as schematically illustrated in **Figure 2f**. Part of these electrons neutralized $V_O^{2+}$s. Thus, a decreased density of $V_O^{2+}$s resulted in an increased Schottky junction width. This explanation is well supported by the strong dependence of the RESET efficiency on the light power density (shown in **Figure 2c**, bottom panel) and on the RESET wavelength (shown in **Figure 2d**). The reason can be understood as follows: irradiation with a relatively short RESET wavelength having relatively high photon energy or with a relatively high light power density generated more electrons injected from metal to ZnO; it follows that more electrons neutralized $V_O^{2+}$s, resulting in a lower density of $V_O^{2+}$s in the Schottky barrier region.

To verify the proposed memconductance tuning mechanism, *i.e.*, optical SET and RESET originated from the respective light-induced narrowing and widening of the Schottky junction, we measured the device capacitance at different irradiation conditions (**Figure 2g**). Capacitance of the device initially increased when irradiated with 530 nm light, but decreased upon a second irradiation with 650 nm light. As mentioned previously, the ZnO/Pt Schottky junction dominated the electrical properties under a positive voltage. This implies that the device capacitance could be controlled by the ZnO/Pt Schottky junction. It has been reported that the capacitance of a Schottky junction increased with decreasing junction width.[65] Hence, we could conclude that the 530 nm irradiation resulted in a narrowing of the Schottky junction (optical SET), whereas the 650 nm irradiation led to a junction widening (optical RESET).

We believe that during the optical SET process, short-wavelength light with relatively high photon energy also induced internal photoemission or photoassisted tunneling of electrons from the metal electrode, and thus causing neutralization of $V_O^{2+}$s. The reason why ionization of $V_O$s dominated during the SET process may be as follows: $V_O$ ionization was apt to occur under short-wavelength irradiation whereas neutralization of $V_O^{2+}$s was limited by the number of electrons injected from metal to the conduction band of ZnO via internal photoemission or photoassisted tunneling; given that the number of injected electrons was severely limited by the Schottky barrier (**Figure 2f**), $V_O^{2+}$ neutralization should play a nondominating role in the optoelectronic response. It deserves mention that the device was driven into a nonequilibrium state where the quasi-Fermi level of electrons of ZnO moved toward the conduction band upon irradiation with short-wavelength light since additional free electrons were generated in ZnO. Due to the difference of Fermi levels between ZnO and metal, the electrons would flow from ZnO to metal spontaneously. These electrons would, to some extent, compensated the electrons injected from metal to ZnO via internal photoemission or photoassisted tunneling, thus leading to further weakened neutralization reaction of $V_O^{2+}$s.

**AOC Memristor**

Based on the optical SET and RESET operations, we can realize the reversible tuning of memconductance by applying only optical excitation. **Figure 3a** exhibits a continuous increase (SET) and decrease (RESET) in the memconductance upon successive light pulses of 530 and 650 nm, respectively. The ON/OFF ratio was calculated to be about 3.2. Herein, ON/OFF ratio was calculated as the ratio of highest memconductance to lowest memconductance. **Figure 3b** shows the results of 20 successive memconductance increase/decrease cycles. The retention behaviors of ten states are illustrated in **Figure 3c**. The memconductance initially showed a slow decay and then remained almost stable. All the states could be clearly distinguished even after $10^4$ s, indicating nonvolatility of the light-induced memconductance states. The decay curves were fitted using exponential functions (**Figure S7**) to further verify their nonvolatility. The fitting results demonstrated that the memconductance could be maintained above a certain value with time, thus confirming the nonvolatile memconductance states. It deserves mentioning that the observed conductance decay is very similar to the PPC effect, an intrinsic phenomenon observed in most optoelectronic devices. Such decay could be explained as follows: the irradiated device was in a nonequilibrium state; electrons in the metal electrode tended to tunnel through the Schottky barrier and were captured by $V_O^{2+}$s, thus resulting in barrier widening. To study the device-to-device variation of the light-induced memconductance states, memconductance increase/decrease cycles from 20 randomly selected devices were measured (**Figure S8**). The lowest and highest memconductance states showed a memconductance range of 95–126 and 310–390 nS, respectively. The ON/OFF ratios were in a range of 2.7–4.1.

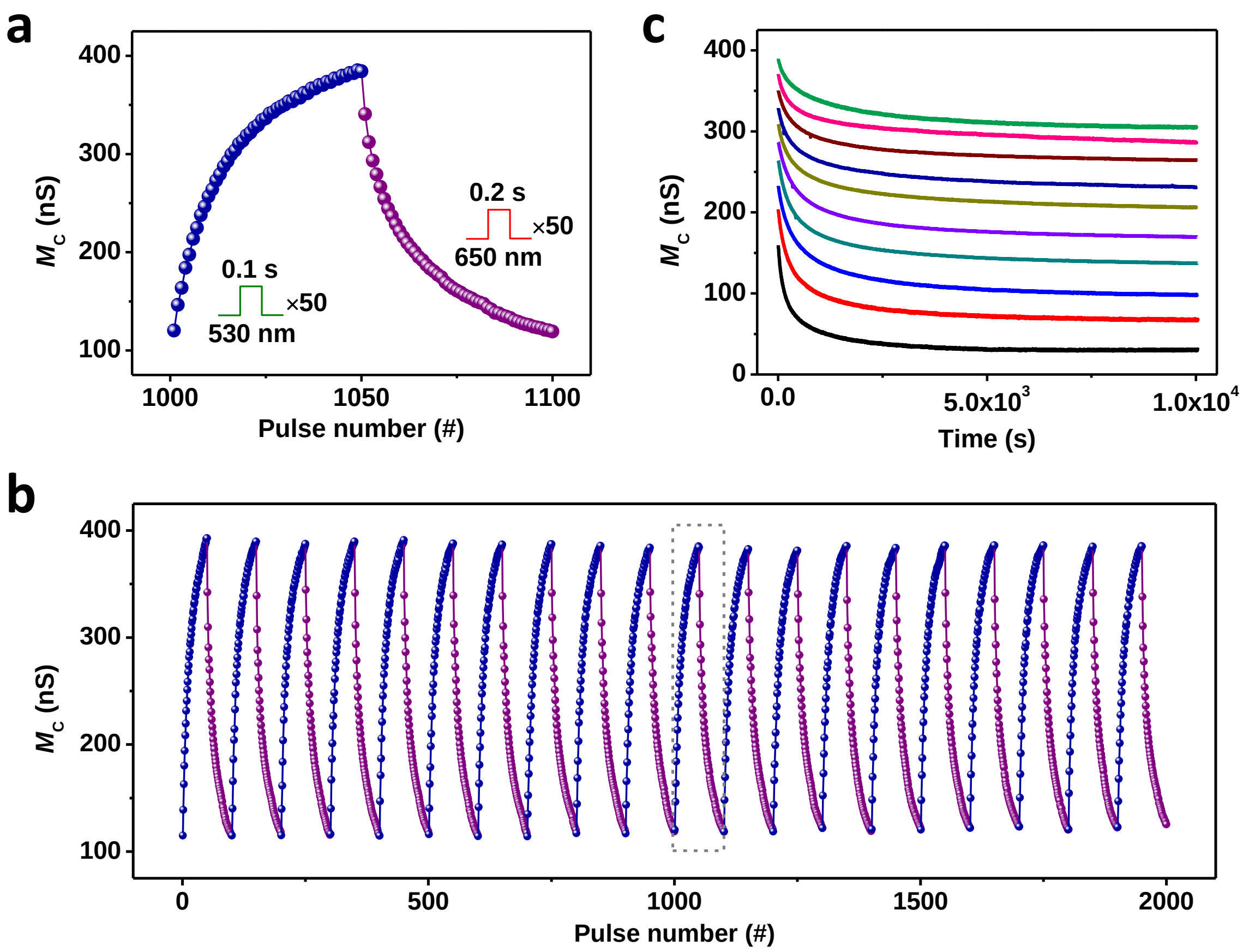

**Figure 3. All-optically tuning of memconductance ($M_C$).** a) Reversible modulation of memconductance by using 50 green light pulses ($D$ = 100 ms and interval ($I$) = 1 s) and 50 red light pulses ($D$ = 200 ms and $I$ = 1 s). b) 20 successive memconductance increase/decrease cycles. An enlarged view of the eleventh cycle (gray rectangle) is shown in (a). c) Retention performance of 10 memconductance states induced by light. In (a) and (b), the memconductance values during the SET and RESET processes were measured just before applying the next pulse, *i.e.*, the delay time between applying light pulse and measuring the memconductance was equal to the interval time of light pulses. All the memconductance values were measured at 10 mV.

To compare the memconductance tuning performance between optically and electrically controlled memristors, we prepared Ti/ZnO/Pt and Cu/ZnO/Pt devices in which the ZnO layer was deposited at the same parameters as the AOC memristor (Au/ZnO/Pt). It has been reported that Ti/ZnO/Pt and Cu/ZnO/Pt demonstrated memristive switching due to electrically controlled electron trapping/detrapping and Cu nanofilament rupture/rejuvenation, respectively.[15,54] After an electroforming process, the Au/ZnO/Pt showed memristive switching based on electrically controlled rupture and rejuvenation of conducting nanofilaments composed of oxygen vacancies.[66] **Figures S9–S11** illustrate electric-induced memristive behaviors of the Ti/ZnO/Pt, Cu/ZnO/Pt and electroformed Au/ZnO/Pt devices, respectively. By comparing the single memconductance increase/decrease cycles in **Figure 3** and **Figures S9–S11**, we could observe that the memconductance increase/decrease curves of the AOC memristor and Ti/ZnO/Pt were much smoother than those of the Cu/ZnO/Pt and electroformed Au/ZnO/Pt. This could be attributed to the purely electronic memristive switching mechanism of the AOC memristor and Ti/ZnO/Pt. Moreover, we found that the cycle-to-cycle and device-to-device variations for the AOC memristor were much smaller than those for the other three devices (**Figure 3** and **S8–S11**). Given that the functional layers of these four memristive devices were deposited at the same parameters, it could be deduced that the AOC memristor has a better memconductance tuning performance than the electrically controlled memristors based on switching mechanisms of both nanofilament rupture/rejuvenation and carrier trapping/detrapping. As for the Ti/ZnO/Pt, the global increase in memconductance during successive cycling likely resulted from the generated Joule heat given the rather high programming voltage and current (**Figure S9a,c**), whereas large variations in the initial memconductance among the devices might be due to various extents to which Ti reacted with ZnO (**Figure S9d**).[15]

We propose that the superior memconductance tuning performance of the AOC memristor could be attributed to its purely electronic memconductance tuning mechanism as well as an extremely small amount of heat generated during the programming process given a very low power density (≈ 30 $\mu$W/cm$^2$) of the programming light. To verify that such weak irradiation did not induce microstructure change in the AOC memristor, we exposed the device to 350 nm light with a power density of 36 $\mu$W/cm$^2$ for a period of time as long as one hour, followed by measuring its memconductance increase/decrease cycles (**Figure S12**). By comparing **Figure 3b** and **Figure S12**, we were unable to observe any obvious performance deterioration after this long-term irradiation. Generally, optoelectronic characteristics of a semiconductor device are extremely sensitive to its microstructure change. Hence, the microstructure change could be excluded in the programming process of this AOC memristor given that the shortest wavelength used in our experiments was 350 nm.

**Nonvolatile Neuromorphic Computing**

Our AOC memristor could be used to perform in-memory computing, *e.g.*, nonvolatile neuromorphic computing. As schematically illustrated in **Figure 4a**, a three-layer artificial neural network (ANN) was constructed with the CrossSim[67] simulator using the measured memconductance values in **Figure 3b** as synaptic weights. The ANN could be utilized to recognize handwritten digits. The following two datasets were employed to train the ANN: small images (8 × 8 pixels) of handwritten digits from the "Optical Recognition of

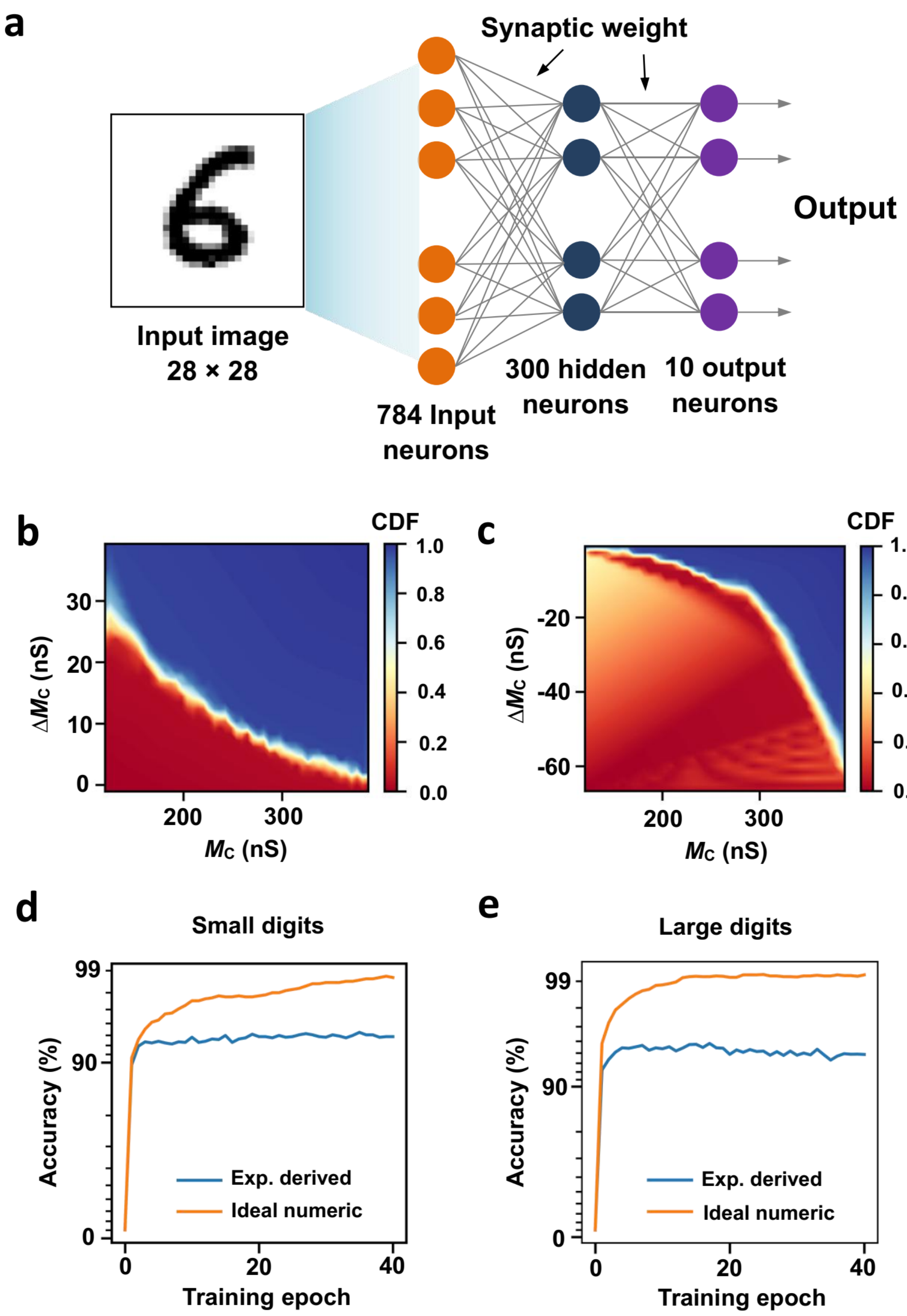

**Figure 4. Image recognition simulations.** a) Schematic illustration of a three-layer ANN for recognition of 28 × 28 pixel images of handwritten digits. b,c) $\Delta M_C$ versus $M_C$ statistics of the AOC memristor for the SET (b) and RESET (c) processes in **Figure 3b**. The heat maps encompass data from 2000 measurements. The color represents the CDF of $\Delta M_C$ at each memconductance state. d,e) Training results based on 8 × 8 pixel images of handwritten digits (d) and 28 × 28 pixel images of handwritten digits (e).

Handwritten Digits" dataset[68] and large images (28 × 28 pixels) of handwritten digits from the "Modified National Institute of Standards and Technology" dataset.[69] During the training process, synaptic weights were updated based on a back-propagation algorithm. **Figure 4b**,**c** shows the cumulative distribution functions (CDFs) for optical SET and optical RESET processes in **Figure 3b**. Herein, CDF is the probability that $\Delta M_C$ takes a value less than or equal to the $\Delta M_C$ plotted. During training, the CDF was randomly sampled for weight update of each synapse. The training results for two datasets are illustrated in **Figure 4d**,**e** (blue curves). We could observe that after three epochs, the recognition accuracy for both small and large images of handwritten digits exceeded 92%. The yellow curves in **Figure 4d**,**e** show the simulation results of the ideal floating-point-based ANN (theoretical limit for the algorithm). The recognition accuracy of the ideal ANN exceeded 98% after 40 epochs, indicating promising application prospects of our AOC memristor in image recognition.

**Logic-in-Memory**

Another type of in-memory computing, logic-in-memory, could also be performed in our AOC memristor. Boolean logic is a form of algebra in which the variable's values are the truth values, *i.e.*, true and false (generally denoted as "1" and "0", respectively). It fits well with the binary numbering system used by modern computers, where each bit has a value of either 1 or 0. There are 16 Boolean logic functions in two-input (*e.g.*, $p$ and $q$) systems. Compared to traditional logic gates based on transistors, logic-in-memory computing architectures provide a more efficient way to store and process information.[10,34–39] Nonvolatile Boolean logic could be demonstrated in our AOC memristor, in which the computing results were *in situ* stored as the memconductance states. As schematically illustrated in **Figure 5a**, one input $p$ is related to the initial memconductance $M_{C0}$ whereas the other input $q$ is related to the light of 530 or 650 nm. The output $p'$ depends on the final memconductance $M_{Cf}$. Except for the OR and NIMP functions, control light of 530 or 650 nm was also needed to demonstrate the other 14 logic functions, *i.e.*, input light and control light have to be applied to the device in certain sequences to obtain required outputs. Herein, 100 nS was set as the baseline value of memconductance; that is, $p$ or $p'$ = 1 meant $M_{C0}$ or $M_{Cf}$ > 100 nS and $p$ or $p'$ = 0 meant $M_{C0}$ or $M_{Cf}$ < 100 nS. $M_{C0}$ was set to be about 140 nS ($p$ = 1) or 70 nS ($p$ = 0). $q$ equals 1 (or 0) when the input light was on (or off). The same rule applied to the control light. In some cases, the control light was applied prior to the input light, referred to as the optoforming operation.

According to the operation complexity, 16 logic functions could be classified in six categories, *i.e.*, i) only needing input light (**Figure 5b** and **Figure S13a**), ii) needing control light (**Figure 5c** and **Figure S13b,c**), iii) needing an optoforming process (**Figure 5d** and **Figure S13d**), iv) needing measurement of $M_{C0}$ (**Figure 5e** and **Figure S13e–h**), v) needing measurement of $M_{C0}$ and optoforming processes (**Figure 5f** and **Figure S13i**), and vi) needing measurement of $M_{C0}$, optoforming processes, and measurement of intermediate memconductance (**Figure 5g** and **Figure S13j**).

Specifically, the OR and NIMP functions could be implemented with or without 530 and 650 nm light irradiation, respectively; the wavelength of input light was determined by the logic operation type. The AND, TRUE, and FALSE functions were demonstrated by sequentially applying the input light and the control light; wavelength of the control light depends on the logic operation type and the input light. For the RIMP and $q$ functions, prior to the input light, the device was exposed to 530 and 650 nm light and set to an HMS and LMS, respectively; wavelength of the control light was selected according to the logic operation type. As for the NAND, IMP, RNIMP, $p$, and NOT $p$ functions, the schemes of applying the control light strongly depend on the initial memconductance state ($p$ = 1 or 0), which should be determined first, as well as the logic operation type and the input light. In the case of NOT $q$ and NOR functions, the optoforming operations were necessary; the schemes of applying the control light were determined by the initial memconductance state and logic operation type. Other than the above 14 logic functions, additional steps were needed to achieve the XOR and NXOR functions. Apart from initial state-dependent optoforming processes, it was necessary to determine the intermediate memconductance states prior to applying the subsequent control light. Furthermore, the control light applied before and after the input light was composed of both 530 and 650 nm irradiation.

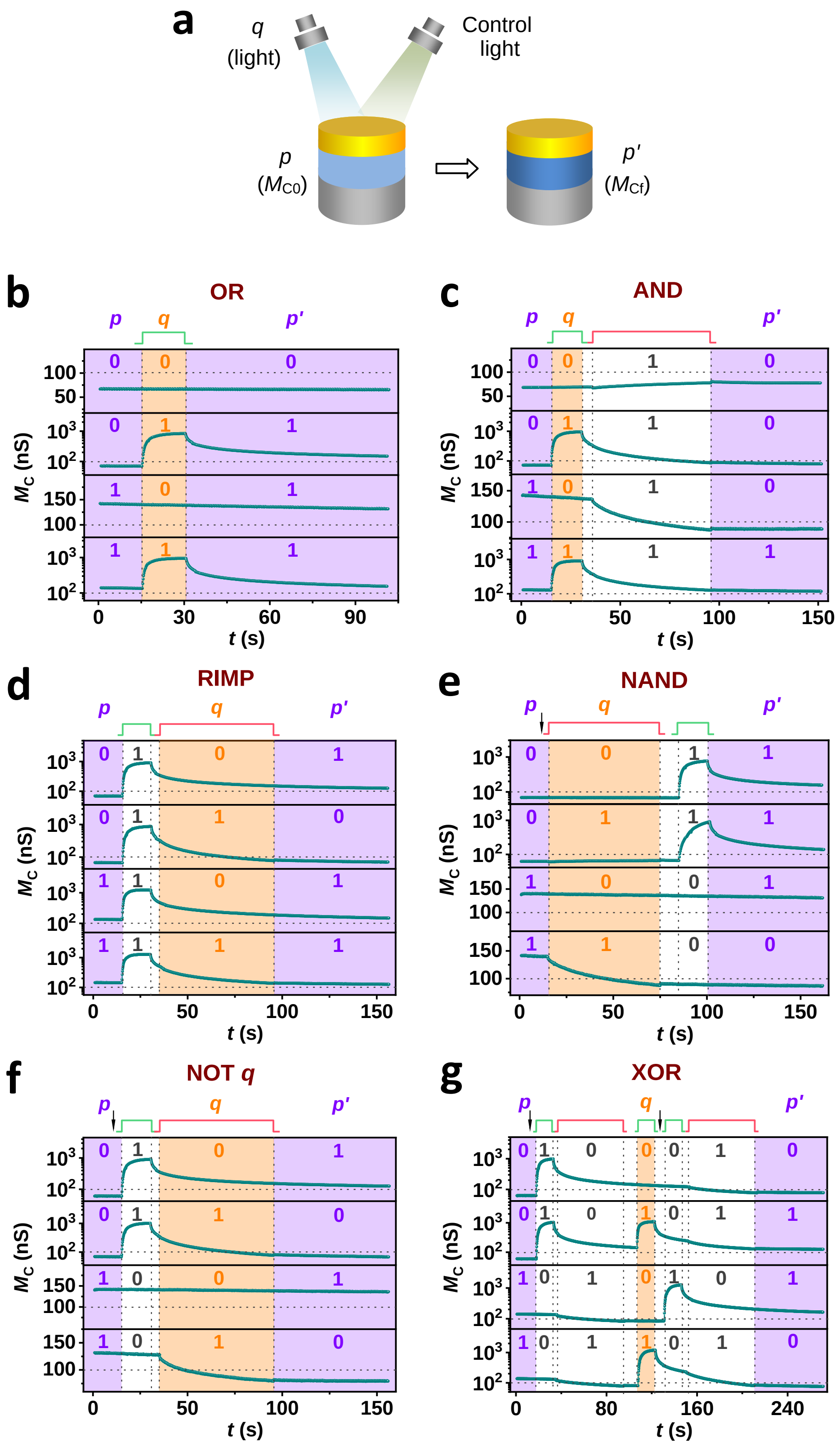

**Figure 5. Nonvolatile logic computing in the AOC memristor.** a) Schematic of the realization of logic functions. *p* and *q* are the inputs and *p′* is the output. $M_{C0}$ and $M_{Cf}$ are the initial and final memconductances, respectively. In most cases, control light is required. b–g) OR, AND, RIMP, NAND, NOT *q*, and XOR logic functions. The green and red polylines represent 530 and 650 nm light, respectively. The black arrows indicate the steps of measuring $M_{C0}$ and intermediate memconductance. The memconductance values were measured at 10 mV.

As mentioned previously, memconductance decay occurs after light irradiation. In the case of $p' = 0$, such decay did not affect the device state since the memconductance gradually deviated away from the baseline value (100 nS). On the other hand, for $p' = 1$, the memconductance gradually approached the baseline value as a result of the decay. To confirm the nonvolatility of the output, the decay curves were fitted using exponential functions (**Figure S14**). The fitting results show that the memconductance could maintain its relatively large value above 100 nS over time, thus indicating the nonvolatile output.

## Conclusions

An AOC memristor with a simple Au/ZnO/Pt structure was fabricated. The memconductance could be reversibly tuned over a continuous range via varying only the wavelength of the controlling light. The device could be operated at light power densities as low as ≈ 30 $\mu$W/cm$^2$ and the light-induced memconductance states were found to be nonvolatile. The observed memconductance switching behavior most likely stemmed from a reversible width variation of the Schottky junction at the metal/ZnO interface due to electron trapping and detrapping at oxygen vacancies. The device showed excellent operation stability. In-memory computing, including nonvolatile neuromorphic computing and logic-in-memory, was demonstrated using this AOC memristor, indicating its great potential as a candidate for high-performance computing.

It deserves mention that before AOC memristors could be practically used for in-memory computing, an answer to the following question must be found, *i.e.*, how can light be controllably introduced into each device to tune its memconductance in a memristor crossbar array? Usage of thin-film optical waveguides based on $Si/SiO_2$, $Si_3N_4/SiO_2$, *etc.*, might be a feasible and effective method to overcome this challenge. These waveguide materials are easy to be integrated into AOC memristors, making high-density integration of AOC memristors possible.

## EXPERIMENTAL PROCEDURES

### Resource Availability

#### Lead Contact

Further information and requests for resources and reagents should be directed to and will be fulfilled by the lead contact, Fei Zhuge (zhugefei@nimte.ac.cn).

#### Materials Availability

This study did not generate any new unique materials.

#### Data and Code Availability

The published article includes all data analyzed during this study.

### Material Growth and Characterization

Polycrystalline ZnO thin films were deposited on $Pt/Ti/SiO_2/Si$, ITO coated $SiO_2/Si$, and quartz substrates at room temperature by RF magnetron sputtering of a ZnO ceramic target of 99.99% purity in pure Ar atmosphere. The sputtering power was 60 W and the pressure was 0.5 Pa. The structure of the as-deposited ZnO films was checked by X-ray diffractometer (XRD, D8 Advance). The thickness was determined to be about 50 nm via variable angle spectroscopic ellipsometry (M-2000 DI, J. A. Woollam Co., Inc.). The resistivity was measured with a Hall effect measurement system (HP-5500C, Nanometrics) using the van der Pauw method. ZnO had a high resistivity, which exceeded the measurement limit (~$10^5$ Ωcm) of the Hall measurement system. Transmittance spectra were measured by a UV–visible–IR spectrophotometer (Lambda 950, PerkinElmer). Photoluminescence spectra were measured via a confocal microscopic Raman spectrometer (Renishaw inVia Reflex, 325 nm). These measurements were performed at room temperature in air.

### Device Fabrication and Characterization

Au top-electrodes with a thickness of 10 nm and a diameter of 100 $\mu$m were deposited onto ZnO films at room temperature by electron beam evaporation with *in situ* metal shadow masks. Ti and Cu top-electrodes with a thickness of 3 and 20 nm were deposited by the same method. To avoid oxidization of Ti and Cu, a 10 nm thick Au protection layer was deopsited onto the Ti and Cu electrodes. ITO top-electrodes with a thickness of 100 nm and a diameter of 100 $\mu$m were deposited at room temperature by RF magnetron sputtering of an ITO ($In_2O_3$:$SnO_2$ = 5:1, molar ratio) ceramic target of 99.99% purity in pure Ar. The sputtering power was 60 W and the pressure was 0.5 Pa. Electrical and optoelectronic measurements were conducted at room temperature in air by a Keithley 4200 semiconductor parameter analyzer equipped with a monochromatic light source (Omni-λ 3007). Voltage was applied to the top electrode (Au or ITO) with the bottom electrode (Pt or ITO) grounded. Light was injected into the device through the top electrode. Unless otherwise specified, the light power density was maintained at 36 $\mu$W/cm$^2$. The fabricated device was in an HMS most likely due to inevitable exposure to light from the environment and the microscope since the device is sensitive to visible light. Such an HMS could be restored to the initial LMS via bias voltage or light irradiation. Unless otherwise specified, the electrical and optoelectronic measurements were based on initialized devices.

## SUPPLEMENTAL INFORMATION

Supplemental information can be found online at http://doi.org/.

## ACKNOWLEDGMENTS

This work is supported in part by the National Natural Science Foundation of China (Nos. U20A20209 and 61874125), the Strategic Priority Research Program of Chinese Academy of Sciences (No. XDB32050204), the Zhejiang Provincial Natural Science Foundation of China (No. LD19E020001) and the State Key Laboratory of Silicon Materials (No. SKL2021-03).

## AUTHOR CONTRIBUTIONS

J.Y., L.H., F.Z. and H.L. conceived and designed the experiments. J.Y., L.H., L.S. and P.C. conducted the experiments. F.Z., H.L., J.W., L.H. and Z.Y. prepared the manuscript. F.Z. and Z.Y. supervised the project. All authors discussed the results and commented on the manuscript.

## DECLARATION OF INTERESTS

The authors declare no competing interests.

Supplemental Information

## In-memory computing based on all-optically controlled memristor

*Jing Yang, Lingxiang Hu, Liufeng Shen, Jingrui Wang, Peihong Cheng, Huanming Lu,* Fei Zhuge* and Zhizhen Ye*

J. Yang,[+] L. X. Hu,[+], L. F. Shen, Dr. J. R. Wang, Prof. P. H. Cheng, Prof. H. M. Lu, Prof. F. Zhuge
Ningbo Institute of Materials Technology and Engineering
Chinese Academy of Sciences
Ningbo 315201, China
E-mail: hmlu@nimte.ac.cn (HL), zhugefei@nimte.ac.cn (FZ)
L. X. Hu, Prof. F. Zhuge
Center of Materials Science and Optoelectronics Engineering
University of Chinese Academy of Sciences
Beijing 100029, China
Prof. F. Zhuge
Center for Excellence in Brain Science and Intelligence Technology
Chinese Academy of Sciences
Shanghai 200072, China
Prof. Z. Z. Ye
State Key Laboratory of Silicon Materials
School of Materials Science and Engineering
Zhejiang University
Hangzhou 310027, China

[+] These authors contributed equally to this work.

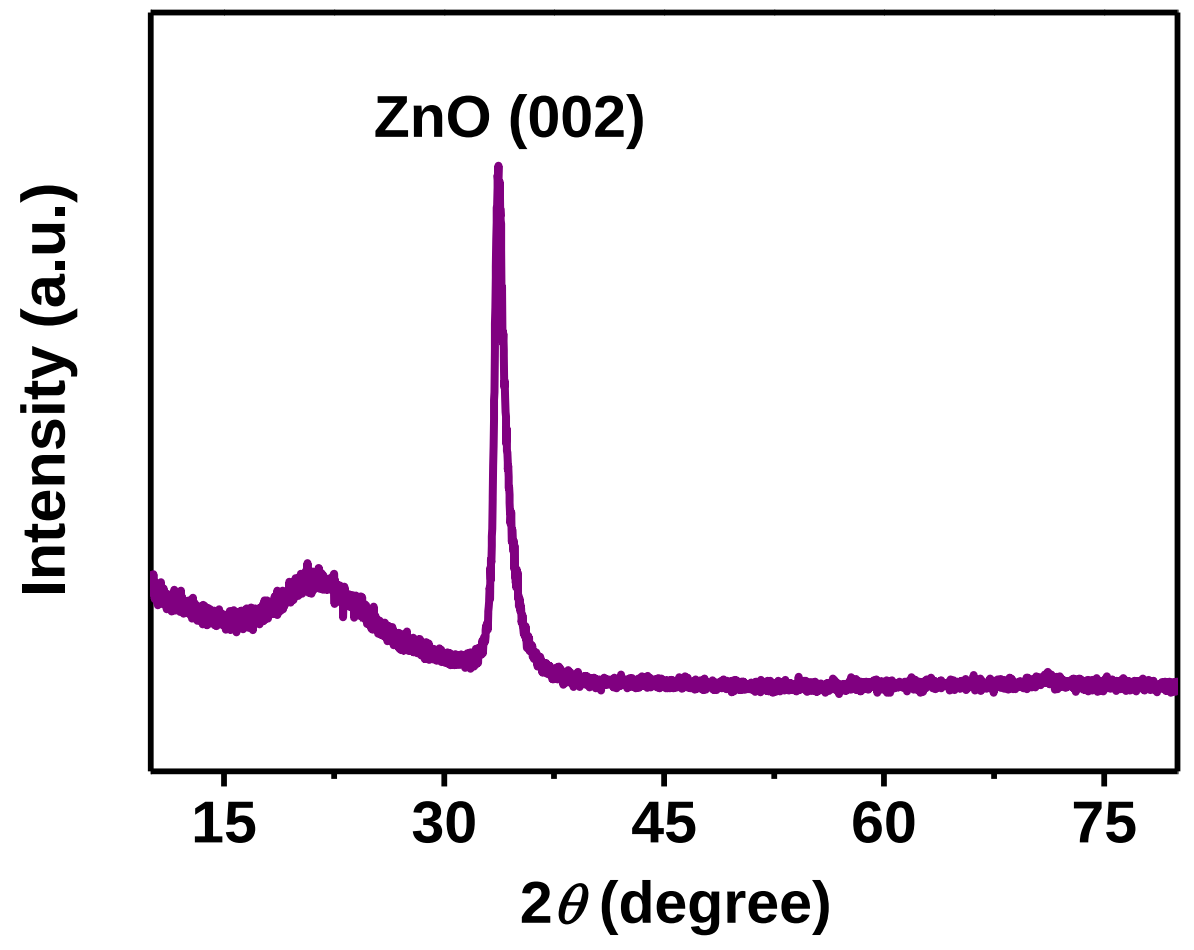

**Figure S1.** XRD pattern of the ZnO thin film deposited on quartz.

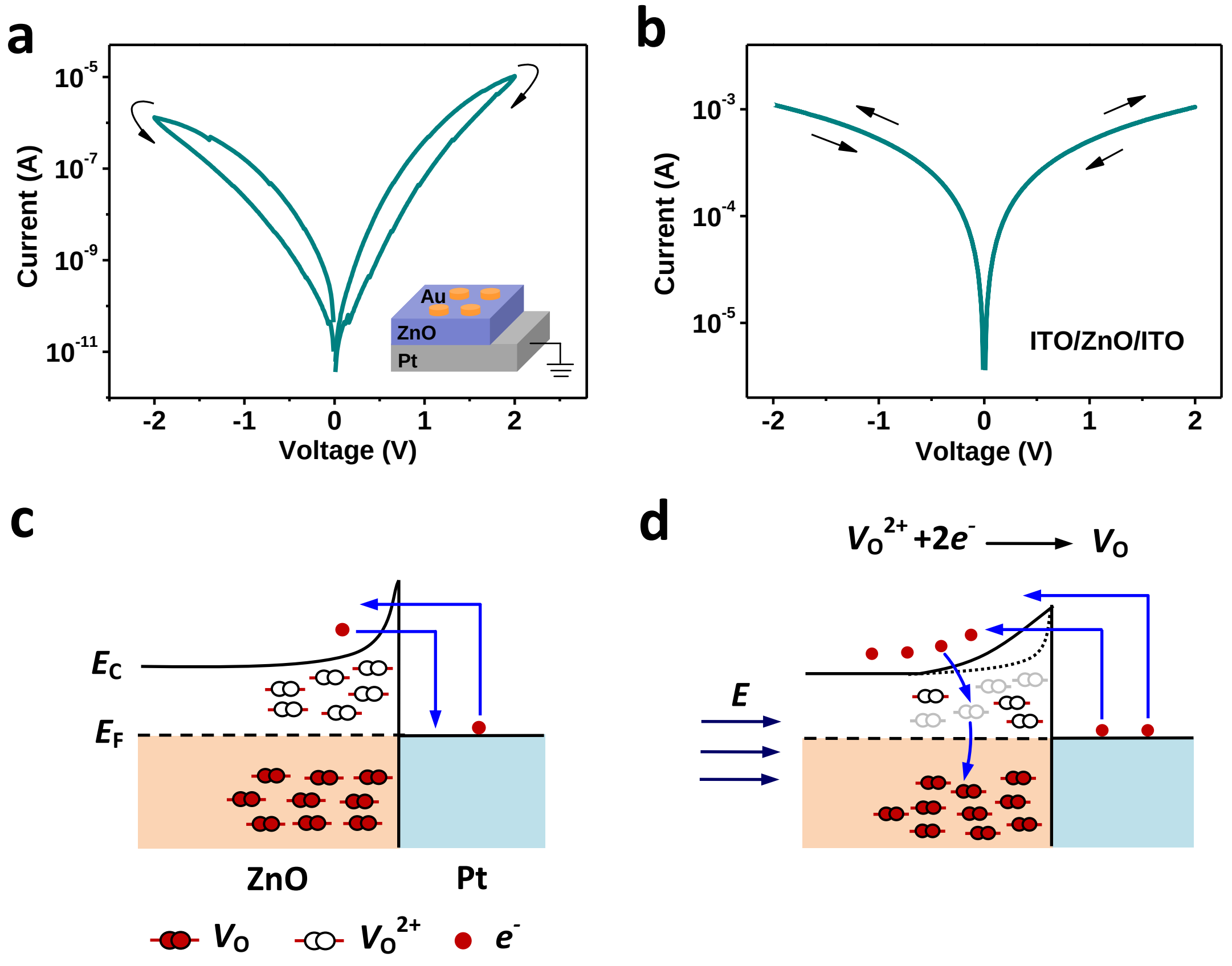

**Figure S2. Nonpolar memristive switching behavior.** a) Current–voltage curves. b) Current–voltage curves of ITO/ZnO/ITO. c) Equilibrium energy band diagram corresponding to the ZnO/Pt Schottky junction in the pristine device. $E_C$ and $E_F$ denote the conduction band minimum and the Fermi energy, respectively. d) Equilibrium energy band diagram after applying a bias voltage or an electric field $E$ (navy arrows). The Schottky emission and tunneling processes of electrons in the Pt electrode as well as the subsequent $V_O^{2+}$

neutralization reaction are also schematically illustrated (blue arrows). The black dotted line indicates the positions of $E_C$ before applying the bias voltage. It deserves mention that both the electron motion and $V_O^{2+}$ neutralization reaction actually occur during nonequilibrium conditions.

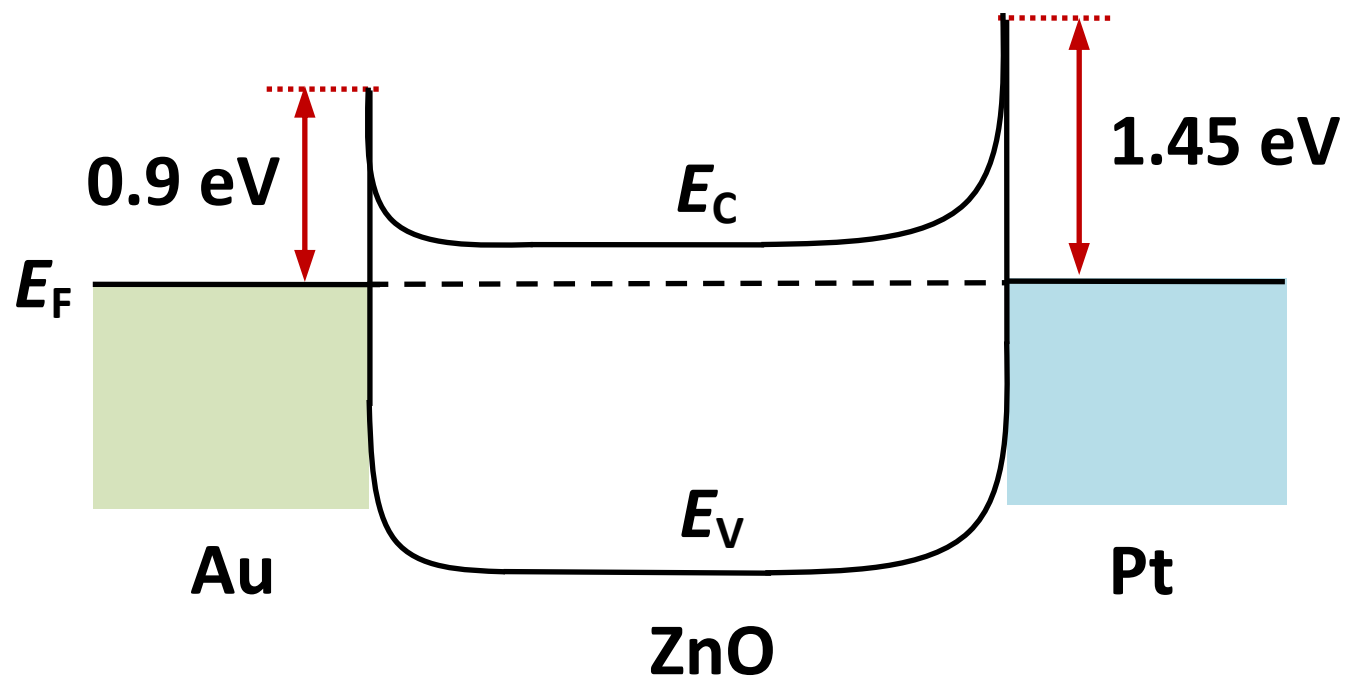

**Figure S3**. Schematic illustration of the Au/ZnO and ZnO/Pt Schottky junctions. The Schottky barrier heights were estimated from the differences between the electron affinity of ZnO (4.2 eV) and the work functions of Au (5.1 eV) and Pt (5.65 eV). $E_F$, $E_C$, and $E_V$ denote the Fermi energy, the conduction band minimum, and the valence band maximum, respectively.

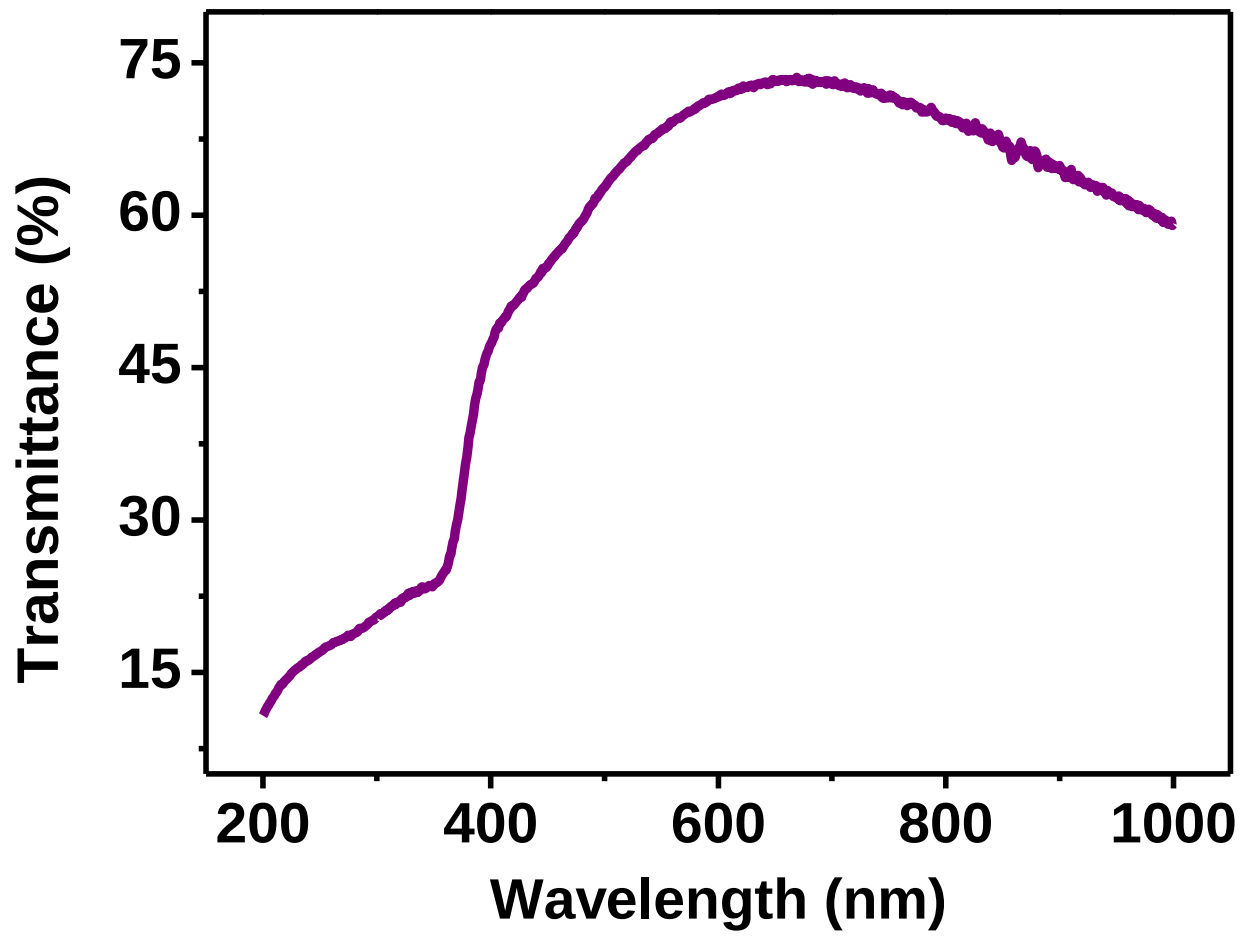

**Figure S4.** Optical transmittance spectrum of Au/ZnO. Thickness of Au is 10 nm.

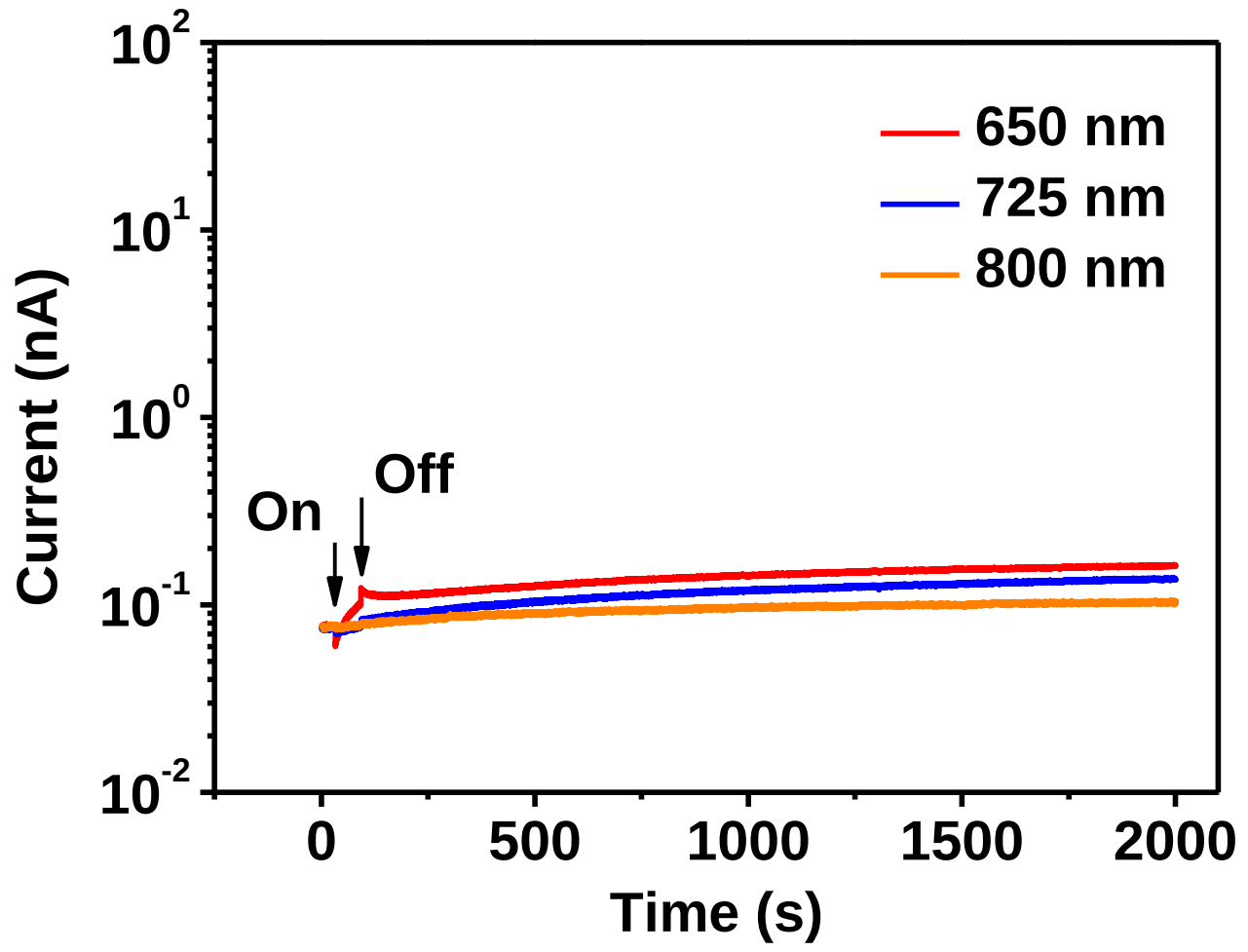

**Figure S5.** Photocurrent responses of the AOC memristor to irradiation with light of 650, 725, and 800 nm. The irradiation duration was 60 s. The current values were measured at 10 mV.

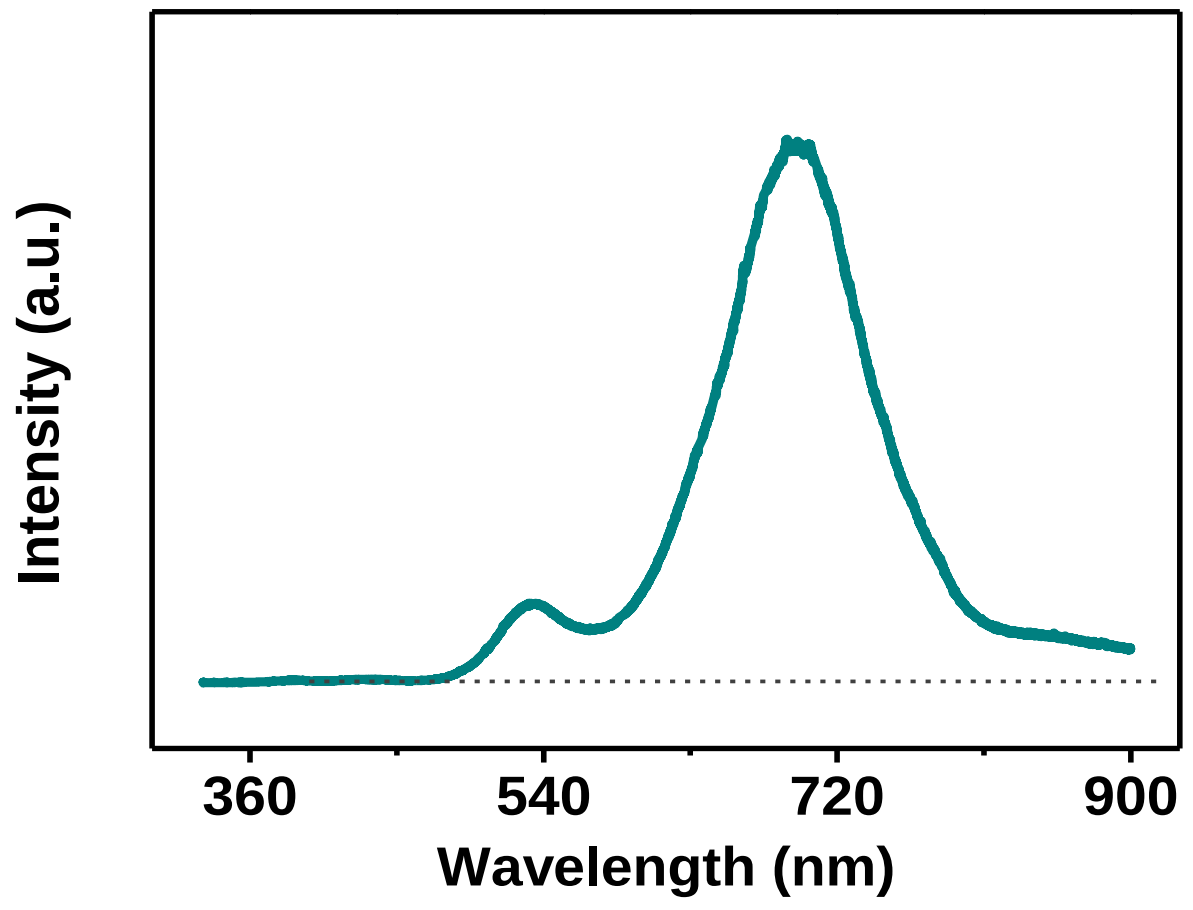

**Figure S6.** Photoluminescence spectra of the ZnO thin film.

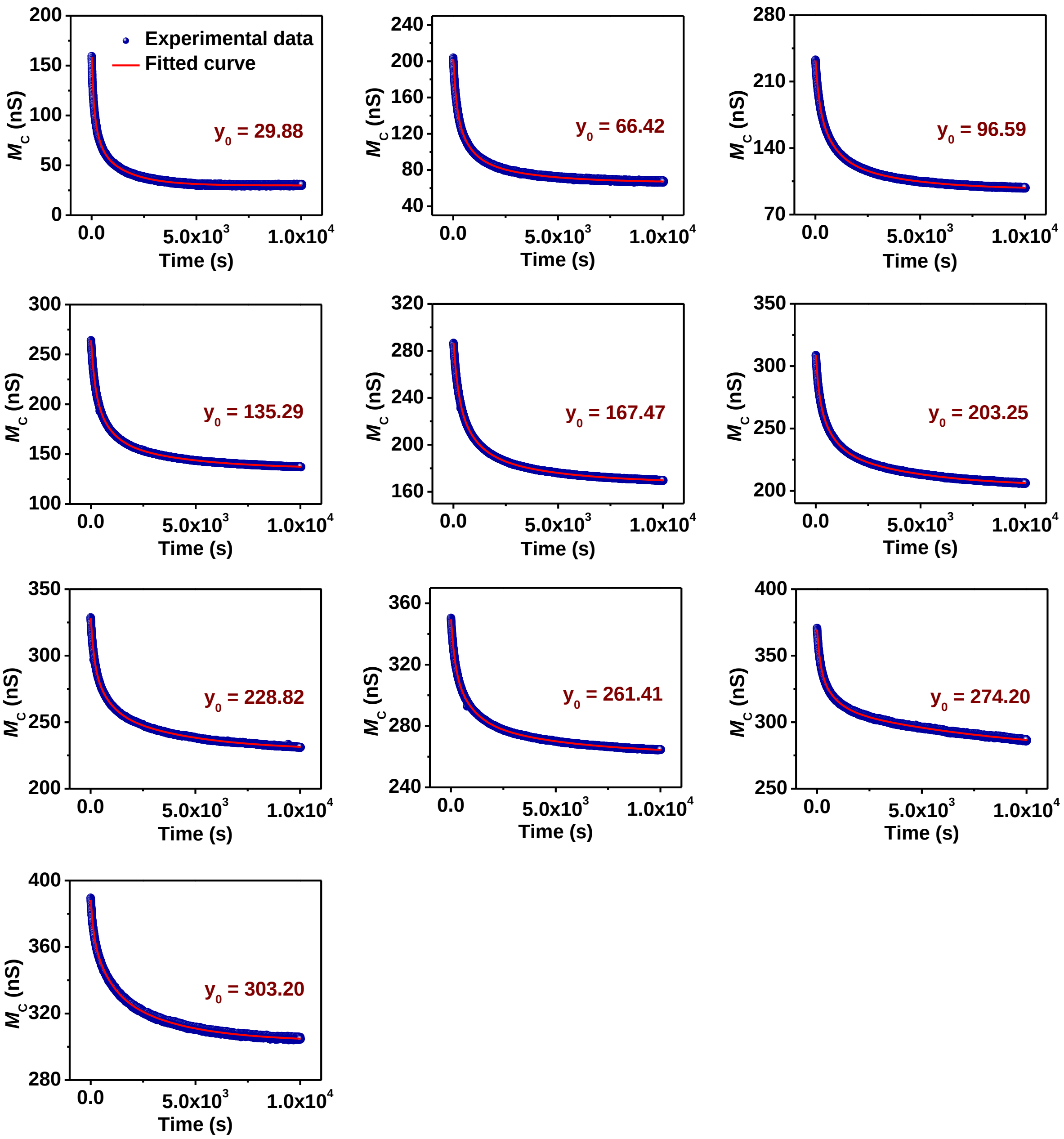

**Figure S7.** Nonvolatility of the 10 memconductance states in **Figure 3c**. The data was fitted by the following exponential function: $M_C = y_0 + A\exp(-t/\tau_1) + B\exp(-t/\tau_2) + C\exp(-t/\tau_3)$, where $y_0$, A, B, C, $\tau_1$, $\tau_2$, and $\tau_3$ are positive constants. The fitting results demonstrate that the memconductance could be maintained above a certain value ($y_0$) over time, thus confirming the nonvolatile memconductance states.

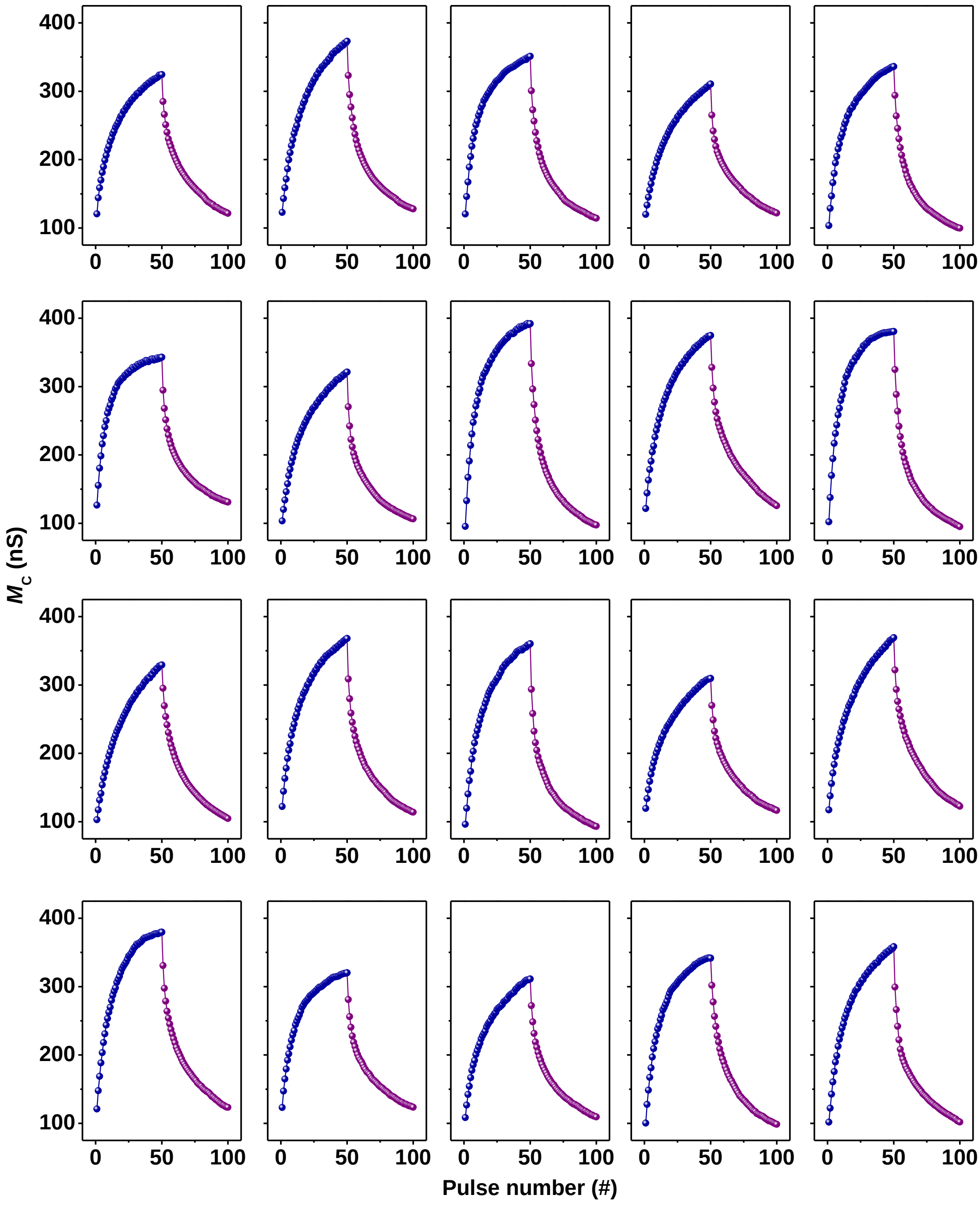

**Figure S8.** Memconductance increase/decrease cycles from 20 randomly selected AOC memristive devices. The reversible modulation of memconductance was carried out by using 50 green light pulses ($D$ = 100 ms and $I$ = 1 s) and 50 red light pulses ($D$ = 200 ms and $I$ = 1 s). The memconductance values were measured 1 s after each pulse at 10 mV.

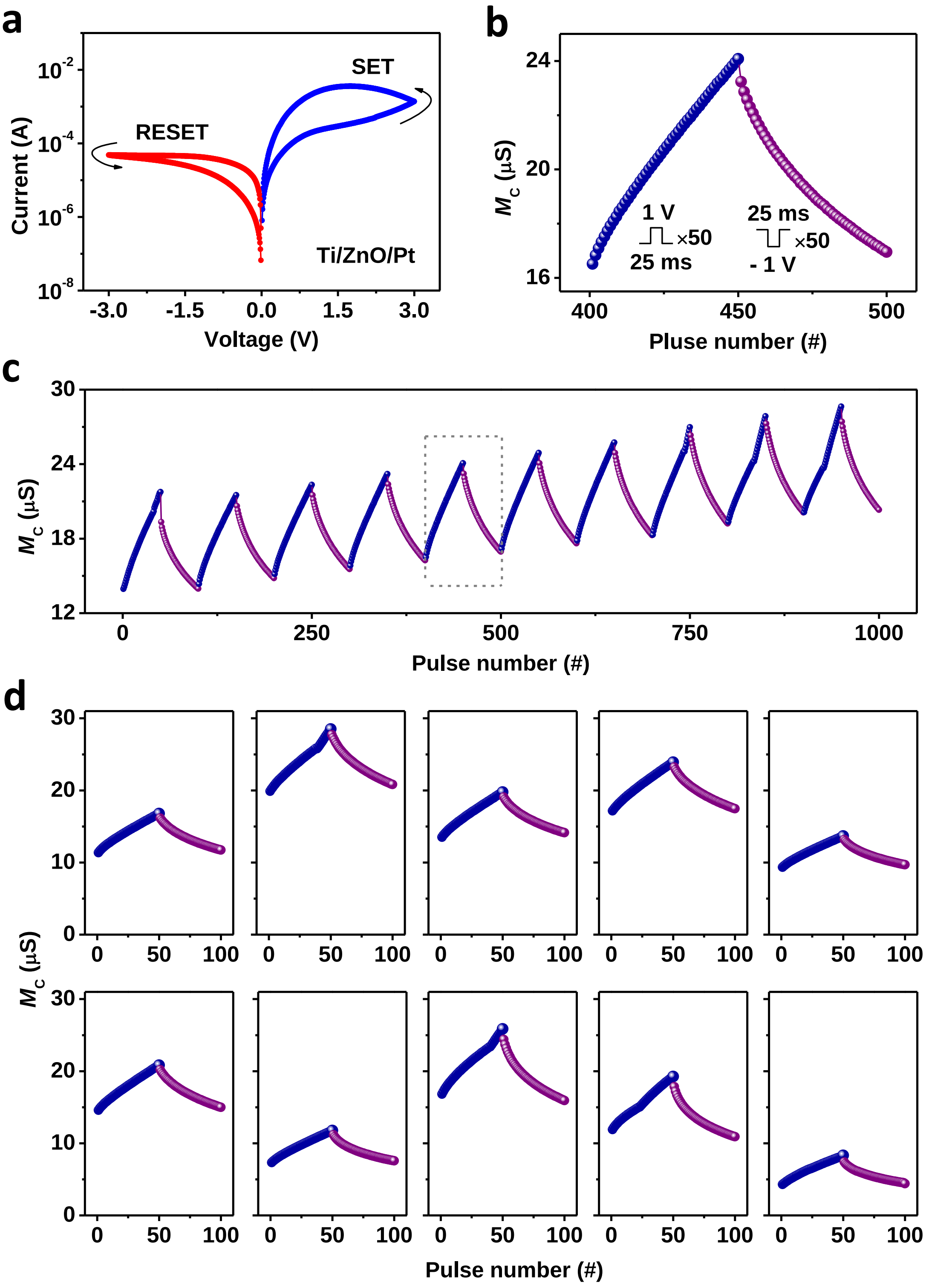

**Figure S9. Ti/ZnO/Pt memristor based on a memristive switching mechanism of electrically controlled electron trapping and detrapping.** a) Current–voltage curves. b) Reversible modulation of memconductance by using 50 positive voltage pulses ($I$ = 100 ms) and 50 negative voltage pulses ($I$ = 100 ms). c) 10 successive memconductance increase/decrease cycles. An enlarged view of the fifth cycle (gray rectangle) is shown in (a). d) Memconductance increase/decrease cycles from 10 randomly selected devices. The memconductance values were measured 100 ms after each pulse at 10 mV.

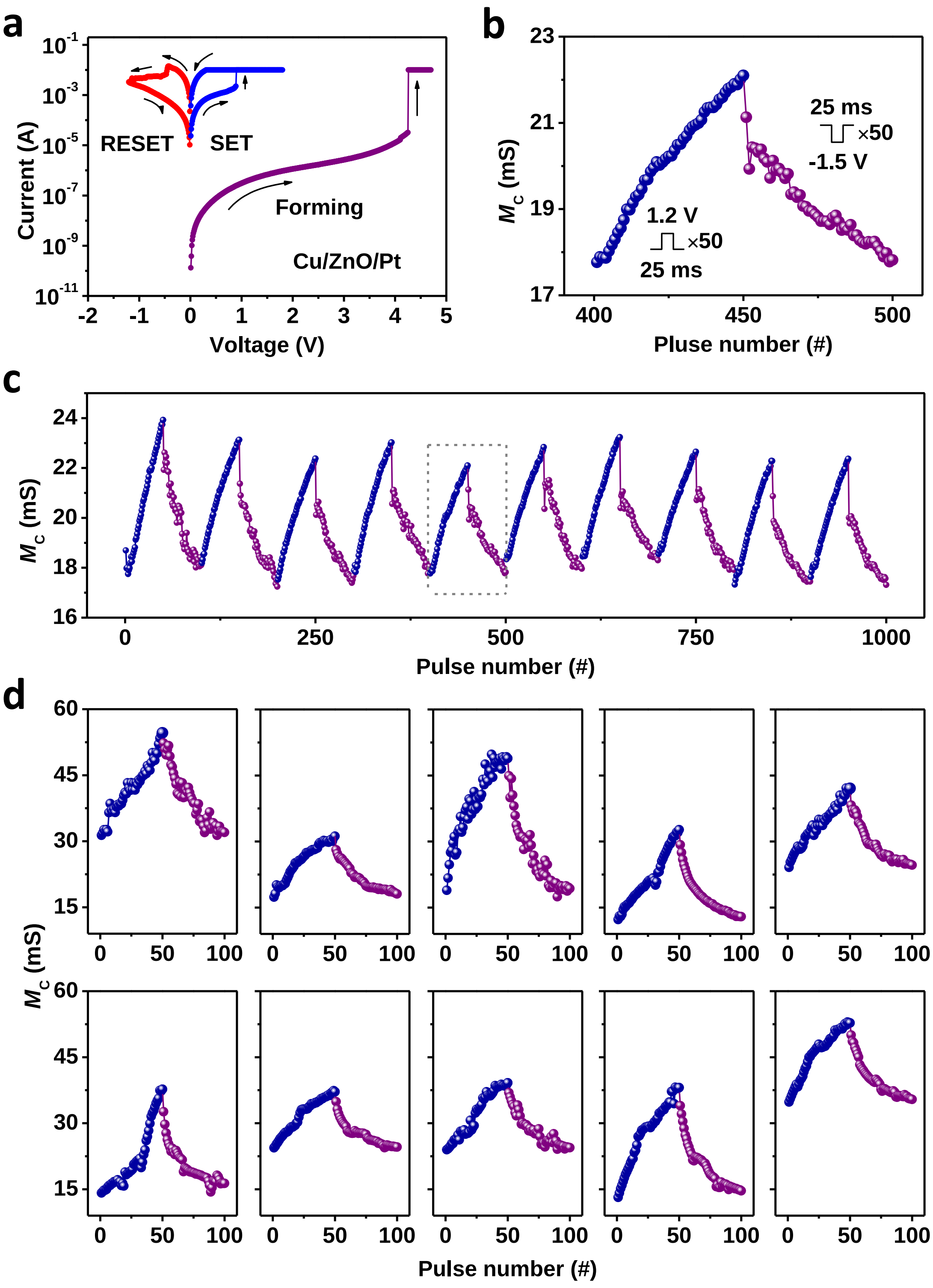

**Figure S10. Cu/ZnO/Pt memristor based on a memristive switching mechanism of electrically controlled Cu nanofilament rupture and rejuvenation.** a) Current–voltage curves. A Forming process was needed to initialize the device. b) Reversible modulation of memconductance by using 50 positive voltage pulses (*l* = 100 ms) and 50 negative voltage pulses (*l* = 100 ms). c) 10 successive memconductance increase/decrease cycles. An enlarged view of the fifth cycle (gray rectangle) is shown in (a). d) Memconductance increase/decrease cycles from 10 randomly selected devices. The memconductance values were measured 100 ms after each pulse at 10 mV.

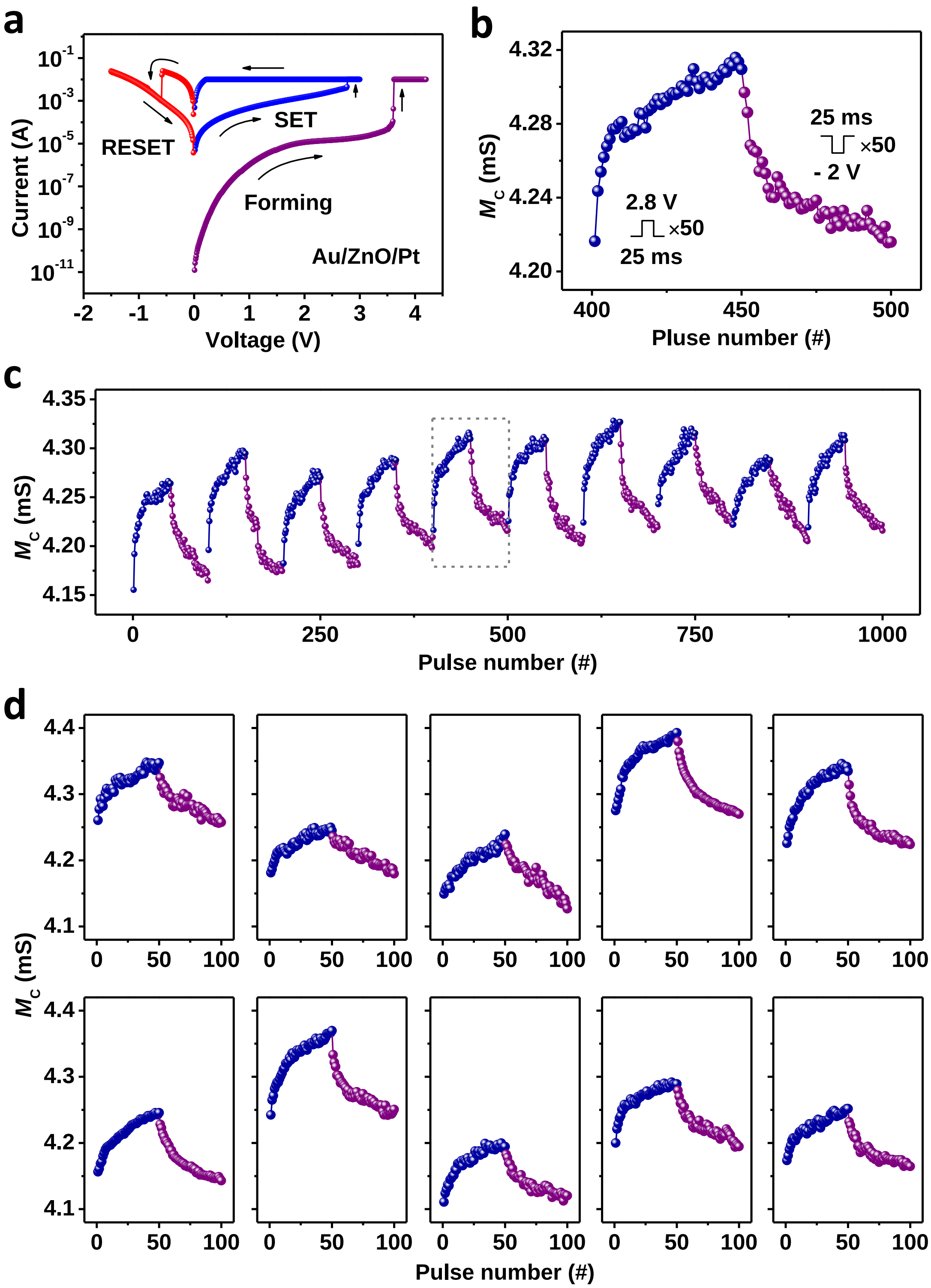

**Figure S11. Au/ZnO/Pt memristor based on a memristive switching mechanism of electrically controlled $V_O$ nanofilament rupture and rejuvenation.** a) Current–voltage curves. A Forming process was needed to initialize the device. b) Reversible modulation of memconductance by using 50 positive voltage pulses ($I$ = 100 ms) and 50 negative voltage pulses ($I$ = 100 ms). c) 10 successive memconductance increase/decrease cycles. An enlarged view of the fifth cycle (gray rectangle) is shown in (a). d) Memconductance increase/decrease cycles from 10 randomly selected devices. The memconductance values were measured 100 ms after each pulse at 10 mV.

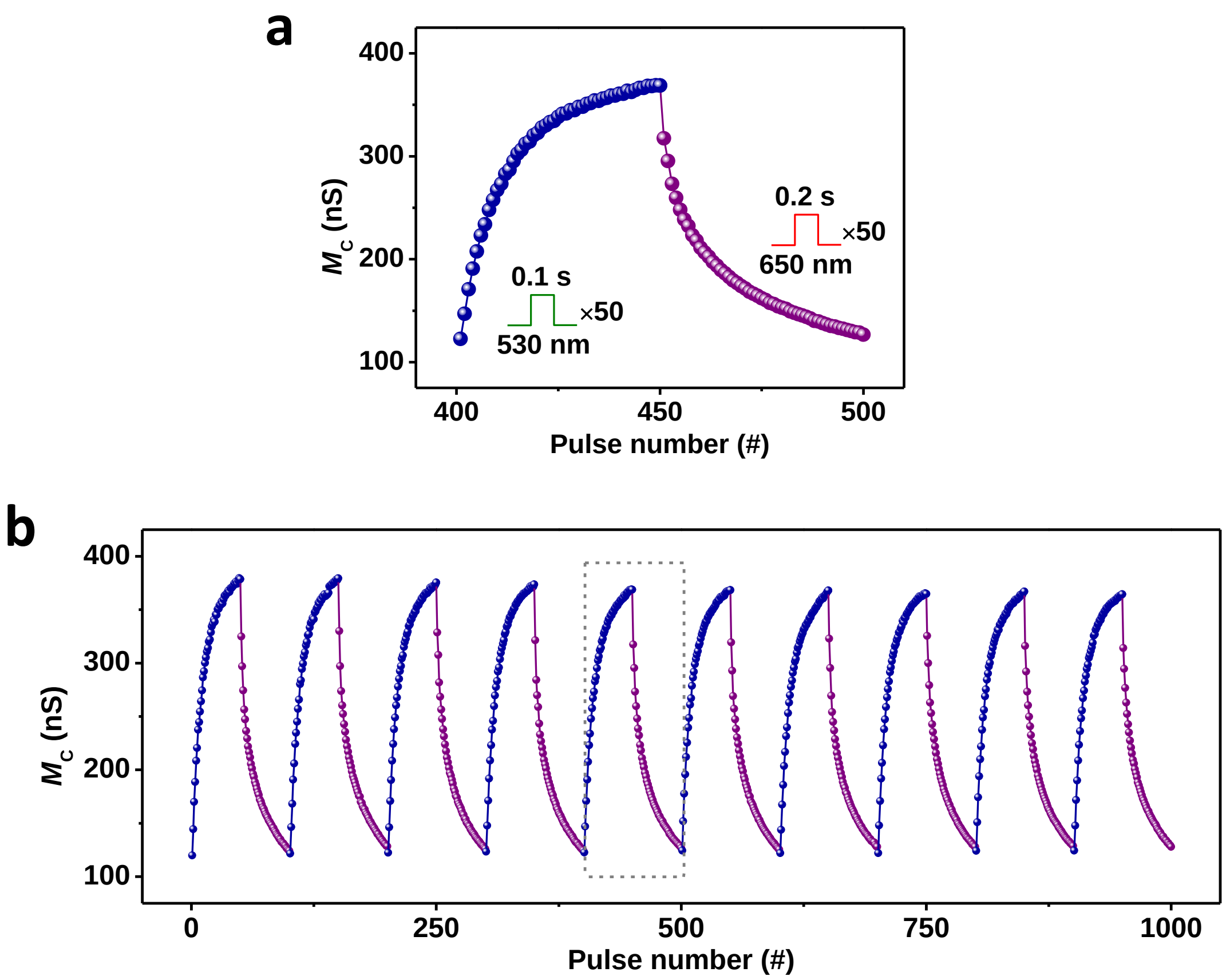

**Figure S12. Memconductance tuning performance of the AOC memristor exposed to 350 nm light with a power density of 36 $\mu W/cm^2$ for one hour.** a) Reversible modulation of memconductance by using 50 green light pulses ($D$ = 100 ms and $I$ = 1 s) and 50 red light pulses ($D$ = 200 ms and $I$ = 1 s). b) 10 successive memconductance increase/decrease cycles. An enlarged view of the fifth cycle (gray rectangle) is shown in (a). The memconductance values were measured 1 s after each pulse at 10 mV.

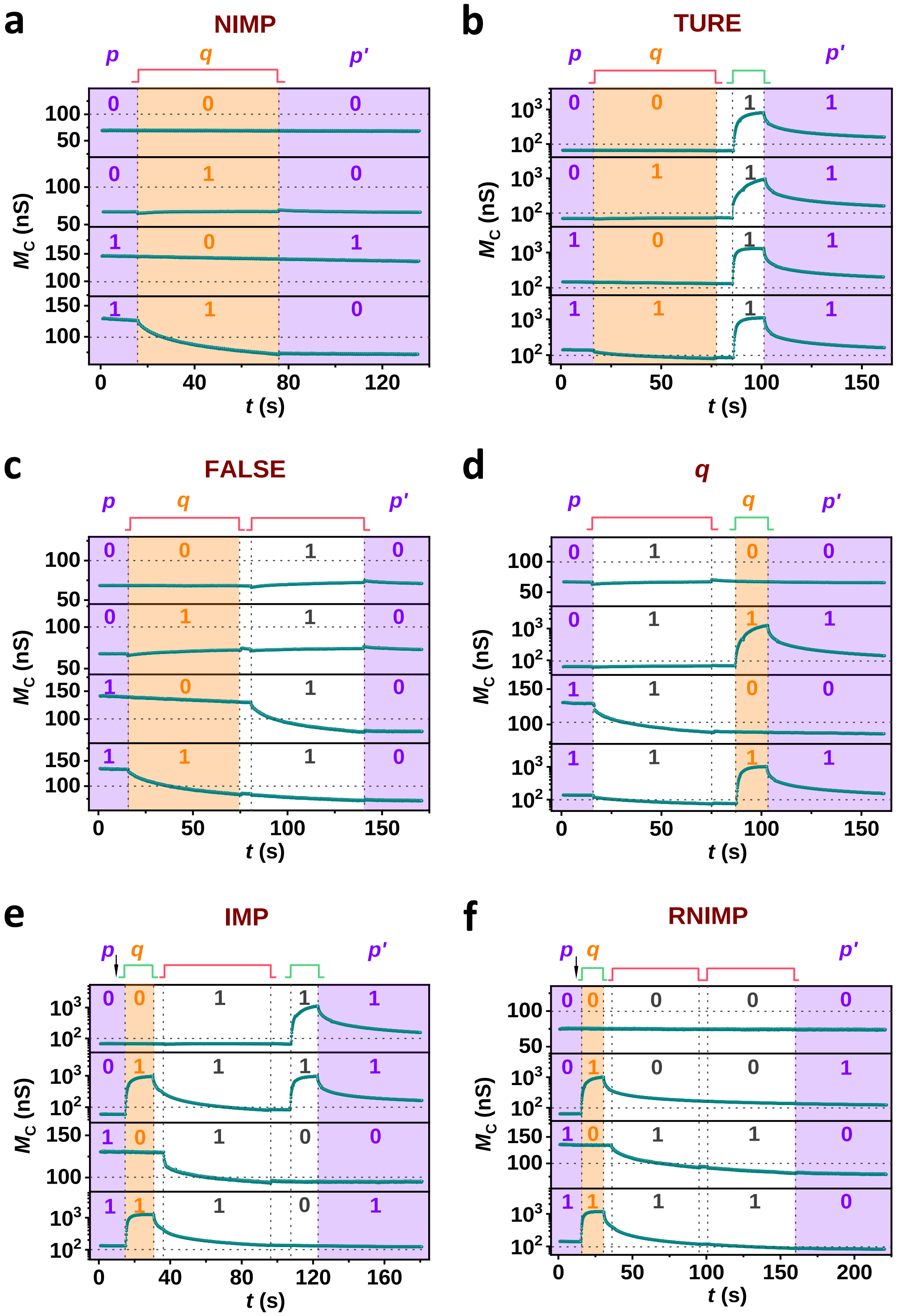

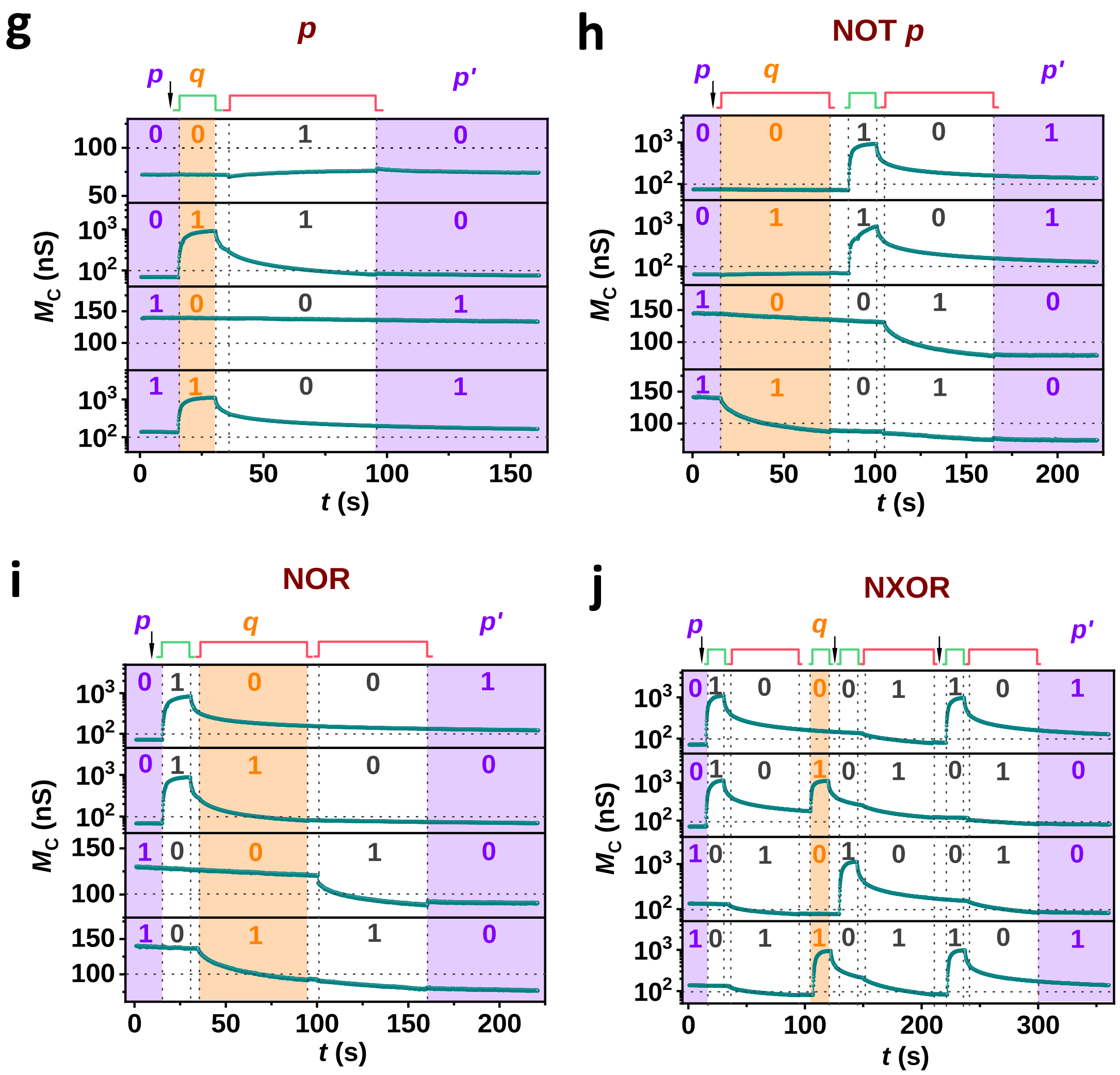

**Figure S13. Nonvolatile logic computing in the AOC memristor.** a–j) NIMP, TRUE, FALSE, *q*, IMP, RNIMP, *p*, NOT *p*, NOR, and NXOR logic functions. The green and red polylines represent 530 and 650 nm lights, respectively. The black arrows indicate the steps of measuring $M_{C0}$ and intermediate memconductance. The memconductance values were measured at 10 mV.

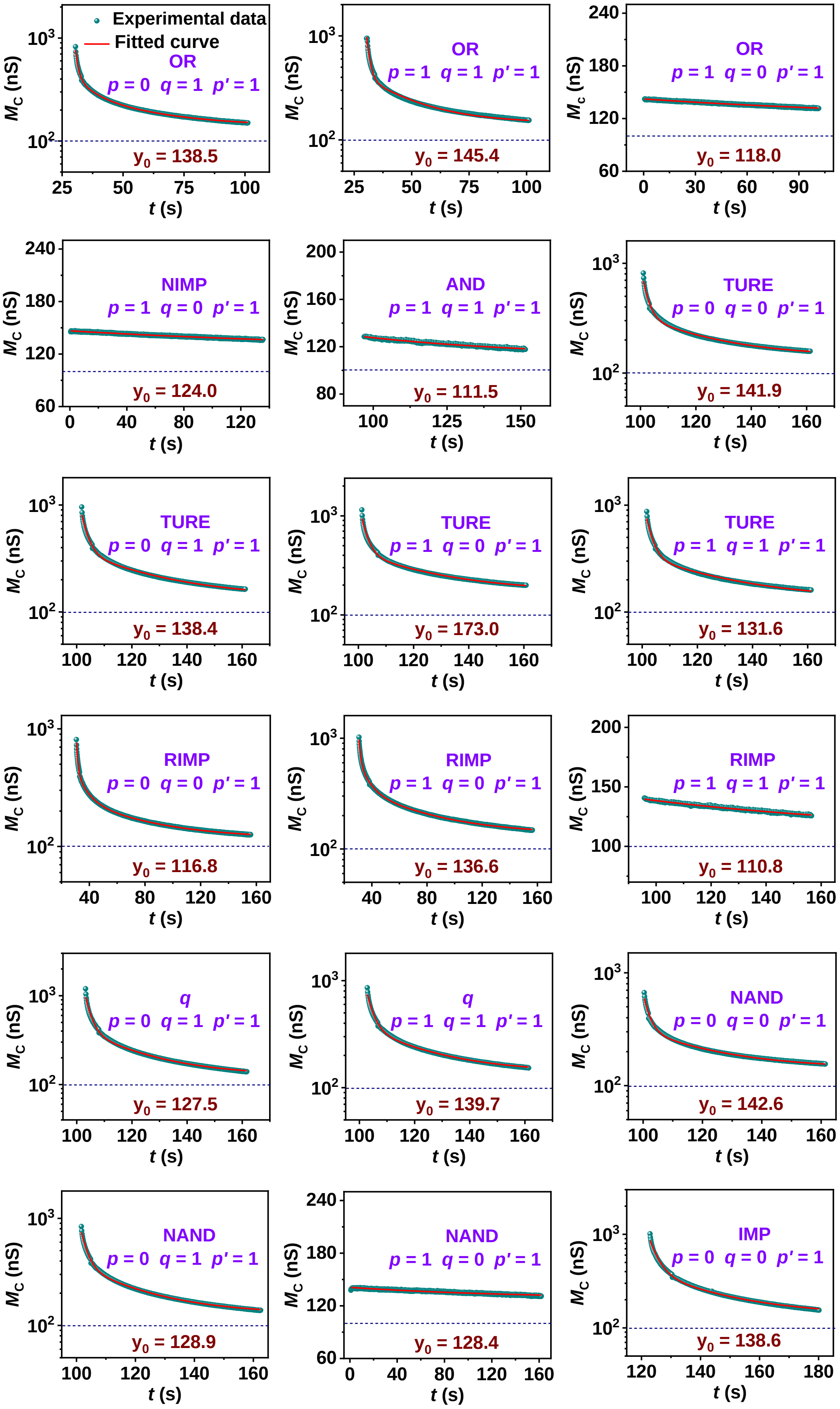
Experimental data
Fitted curve
$M_C$ (nS)
t (s)
OR
p = 0 q = 1 p′ = 1
y0 = 138.5
OR
p = 1 q = 1 p′ = 1
y0 = 145.4
OR
p = 1 q = 0 p′ = 1
y0 = 118.0
NIMP
p = 1 q = 0 p′ = 1
y0 = 124.0
AND
p = 1 q = 1 p′ = 1
y0 = 111.5
TURE
p = 0 q = 0 p′ = 1
y0 = 141.9
TURE
p = 0 q = 1 p′ = 1
y0 = 138.4
TURE
p = 1 q = 0 p′ = 1
y0 = 173.0
TURE
p = 1 q = 1 p′ = 1
y0 = 131.6
RIMP
p = 0 q = 0 p′ = 1
y0 = 116.8
RIMP
p = 1 q = 0 p′ = 1
y0 = 136.6
RIMP
p = 1 q = 1 p′ = 1
y0 = 110.8
q
p = 0 q = 1 p′ = 1
y0 = 127.5
q
p = 1 q = 1 p′ = 1
y0 = 139.7
NAND
p = 0 q = 0 p′ = 1
y0 = 142.6
NAND
p = 0 q = 1 p′ = 1
y0 = 128.9
NAND
p = 1 q = 0 p′ = 1
y0 = 128.4
IMP
p = 0 q = 0 p′ = 1
y0 = 138.6

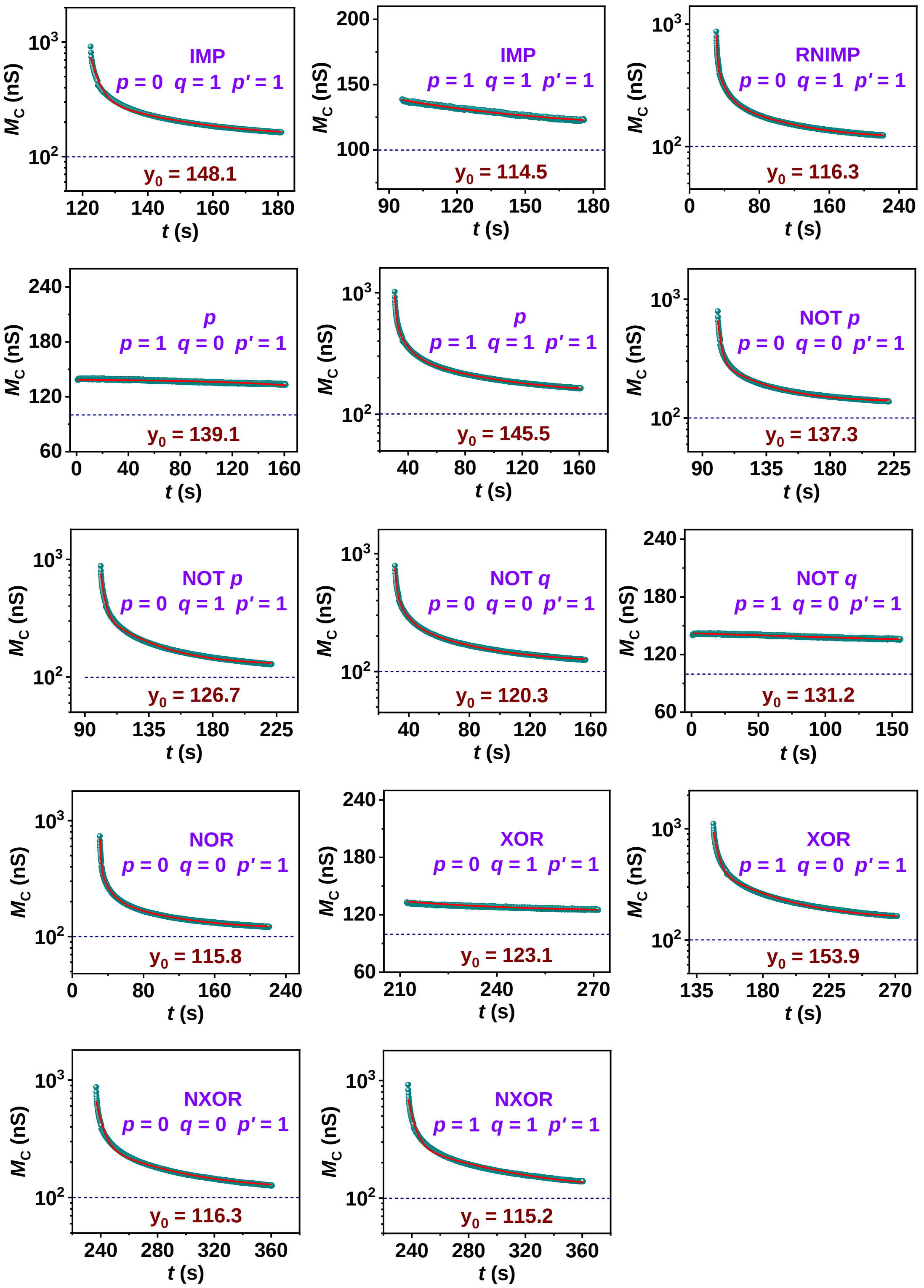

**Figure S14.** Nonvolatility of the output ($p'$ = 1) of the logic operations in **Figure 5** and **Figure S13**. The data was fitted by the following exponential function: $M_C = y_0 + A\exp(-t/\tau_1) + B\exp(-t/\tau_2) + C\exp(-t/\tau_3)$, where $y_0$, A, B, C, $\tau_1$, $\tau_2$ and $\tau_3$ are positive constants. The fitting results demonstrate that a relatively large value of the memconductance (above 100 nS) could be maintained over time, thus confirming the nonvolatile output.